\documentclass{aa}
\usepackage[varg]{txfonts}

\usepackage{graphicx}
\usepackage{natbib}
\bibpunct{(}{)}{;}{a}{}{,} 
\usepackage{color}
\usepackage{float}
\usepackage{subfigure}

\newcommand{\rev}[1]{{#1}}

\usepackage{enumitem}
\setlist{nolistsep}

\begin{document}

\title{The MUSE Hubble Ultra Deep Field Survey}
\subtitle{VI: The Faint-End of the Ly$\alpha$ Luminosity Function at 
  $2.91 < z < 6.64$\\
 and Implications for Reionisation}

\author{A. B. Drake\inst{\ref{inst1}}
\and T. Garel\inst{\ref{inst1}}
\and {\rev{L. Wisotzki\inst{\ref{inst2}}}}
\and F. Leclercq\inst{\ref{inst1}}
\and T. Hashimoto\inst{\ref{inst1}}
\and J. Richard\inst{\ref{inst1}}
\and R. Bacon\inst{\ref{inst1}}
\and J. Blaizot\inst{\ref{inst1}}
\and J. Caruana\inst{\ref{inst3},\ref{inst4}}
\and S.Conseil\inst{\ref{inst1}}
\and T. Contini\inst{\ref{inst5}} 
\and B. Guiderdoni\inst{\ref{inst1}}
\and E. C. Herenz \inst{\ref{inst2}}
\and H. Inami\inst{\ref{inst1}}
\and J. Lewis\inst{\ref{inst1}}
\and G. Mahler\inst{\ref{inst1}}
\and R. A. Marino\inst{\ref{inst7}}
\and R. Pello \inst{\ref{inst5}}
\and J. Schaye\inst{\ref{inst6}} 
\and A. Verhamme\inst{\ref{inst8},\ref{inst1}}
\and E. Ventou\inst{\ref{inst5}} 
\and P. M. Weilbacher\inst{\ref{inst2}}
}

\institute{Univ Lyon, Univ Lyon1, Ens de Lyon, CNRS, Centre de Recherche Astrophysique de Lyon UMR5574, F-69230, Saint-Genis-Laval, France\\\label{inst1}
\email{alyssa-bryony.drake@univ-lyon1.fr}\label{email}
\and Leibniz-Institut fur Astrophysik Potsdam (AIP), An der Sternwarte
16, 14482 Potsdam, Germany\label{inst2}
\and Department of Physics, University of Malta, Msida MSD 2080, Malta. \label{inst3}
\and Institute of Space Sciences \& Astronomy, University of Malta,
Msida MSD 2080, Malta \label{inst4}
\and Institut de Recherche en Astrophysique et Planétologie (IRAP), Université de Toulouse, CNRS, UPS, F-31400 Toulouse, France\label{inst5}
\and Leiden Observatory, P.O. Box 9513, NL-2300 RA Leiden, The Netherlands \label{inst6}
\and Institute for Astronomy, ETH Zurich, Wolfgang-Pauli-Strasse 27,
8093 Zurich, Switzerland \label{inst7}
\and Observatoire de Geneve, Universite de Geneve, 51 Ch. des
Maillettes, 1290 Versoix, Switzerland \label{inst8}
\\
}


\abstract
{
We present the deepest study to date of the Ly$\alpha$ luminosity
function in a blank field using blind integral field spectroscopy from MUSE. We constructed a sample of $604$ Ly$\alpha$ emitters (LAEs) across the redshift range $2.91 < z <
6.64$ using automatic detection software in the Hubble Ultra Deep Field. The deep data cubes
     allowed us to calculate accurate total Ly$\alpha$ fluxes capturing
     low surface-brightness extended Ly$\alpha$ emission now known to be a
     generic property of high-redshift star-forming galaxies. We simulated realistic
     extended LAEs to fully characterise the selection function of our
     samples, and performed flux-recovery experiments to test and 
     correct for bias in our determination of total Ly$\alpha$
     fluxes. We find that an accurate completeness correction accounting
     for extended emission reveals a very steep faint-end slope of the luminosity
   function, $\alpha$, down to luminosities of  log$_{10}$
     L erg s$^{-1}$$< 41.5$, applying both the $1/V_{max}$ and maximum likelihood
   estimators. Splitting the sample into three broad redshift bins, we
   see the faint-end slope
   increasing from $-2.03^{+ 1.42}_{-0.07}$  at  $z \approx 3.44$ to $-2.86^{+0.76}_{-\infty}$ at $z
   \approx 5.48$, however no strong evolution is seen between the
$68\%$ confidence regions in $L^*$-$\alpha$ parameter space. Using the Ly$\alpha$ line flux as a proxy for
   star formation activity, and integrating the observed luminosity
   functions, we find that LAEs' contribution to the cosmic star
   formation rate density
         rises with redshift until it is comparable to that from continuum-selected samples by $z \approx 6$. This implies that LAEs
     may contribute more to the star-formation activity of the early
     Universe than previously thought, as any additional inter-glactic
     medium (IGM)
     correction would act to further boost the Ly$\alpha$
     luminosities. Finally, assuming fiducial
     values for the escape of Ly$\alpha$ and LyC radiation, and the
     clumpiness of the IGM, we integrated the maximum
     likelihood luminosity function at {\mbox{$5.00 < z <
     6.64$}} and find we require only a small
  extrapolation beyond the data ($< 1$ dex in luminosity) for LAEs alone to 
     maintain an ionised IGM at $z \approx 6$.
}


\maketitle

\section{Introduction}

The epoch of reionisation represents the last dramatic phase change of
the Universe, as the neutral intergalactic medium was transformed
by the first generation of luminous objects to the largely ionised
state in which we see it today. We now know that reionisation was
complete by $z \approx 6$, however little is known about the nature of
the sources that powered the reionisation process.

 In recent years, studies have turned towards assessing the number densities and
ionising power of different classes of objects. While the hard ionising
spectra of quasars made them prime candidates, their number densities
proved not to be great enough to produce the ionising photons required
(\citealt{Jiang08}, \citealt{Willott05}). Thus, attention turned to
whether {\mbox{`normal'}} star-forming galaxies were the main drivers of this
process. Traditionally, broadband-selected rest-frame UV samples are used to assess the number
density of star-forming galaxies (\citealt{Bunker04}, 2010, \citealt{Schenker12}, 2013,
\citealt{Bouwens15a}). These studies revealed two very important things: that low-mass galaxies dominate
the luminosity budget at high redshift, and that only $\sim 18\%$ of
these galaxies can be detected via their UV emission alone using
current facilities (\citealt{Bouwens15b}, \citealt{Atek2015}).

\begin{figure*}
\begin{center}
\includegraphics[width=0.48\textwidth]{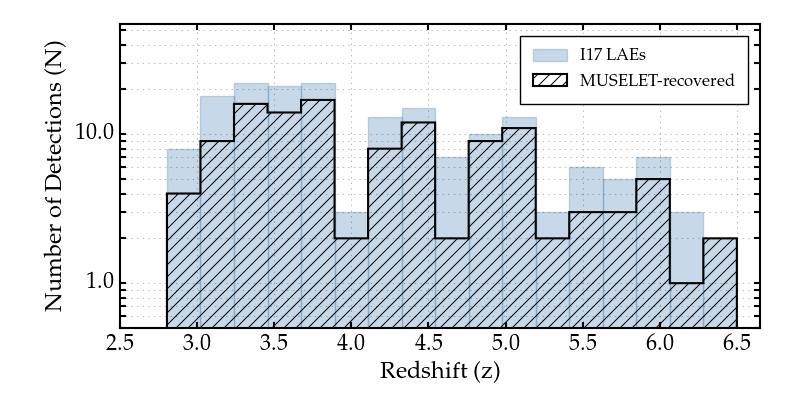} 
 \includegraphics[width=0.48\textwidth]{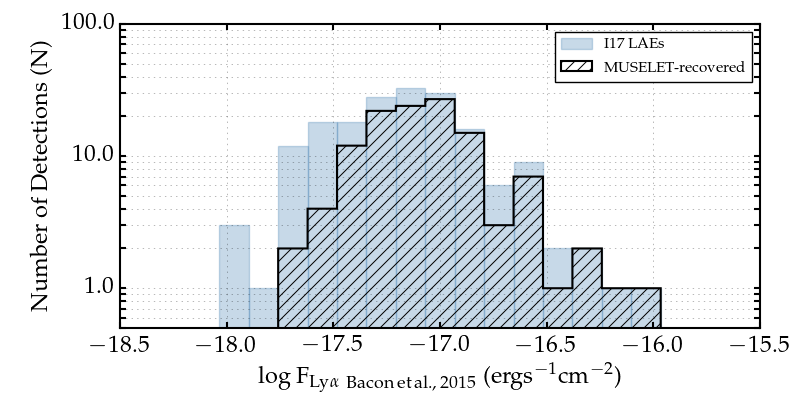} 
 \includegraphics[width=0.48\textwidth]{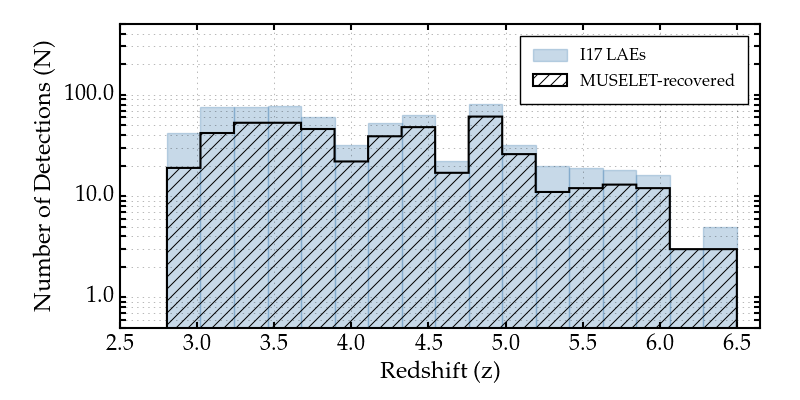} 
 \includegraphics[width=0.48\textwidth]{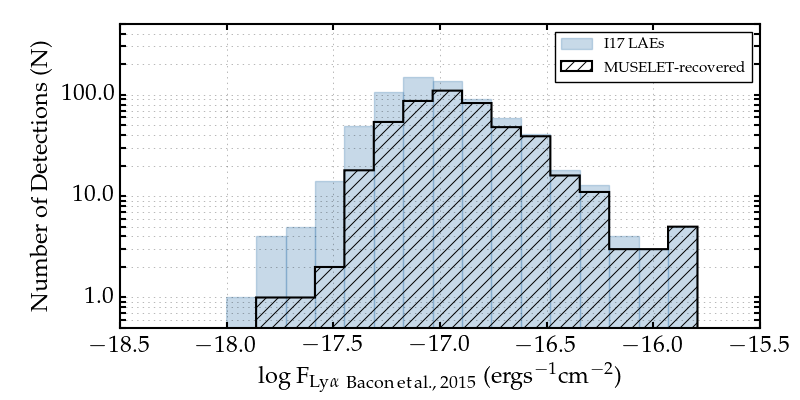} 
\caption[]{Detections using our detection software
   {\sc{muselet}}, overplotted on the LAEs discovered and catalogued
   in I17. In the left-hand panels (upper {\emph{udf-10}}, lower {\emph{mosaic}}) we show the redshift distributions,
   demonstrating an even recovery rate across the entire redshift
   range. In the
   right-hand panels (upper {\emph{udf-10}}, lower {\emph{mosaic}}) we use the published flux estimates of I17 to show the distribution of fluxes
   recovered by {\sc{muselet}} vs the distribution for I17 LAEs. }
\label{fig:nocounts}
\end{center}
\end{figure*}

Detecting galaxies by virtue of their Ly$\alpha$ emission however,
gives us access to the low-mass end of the star-forming galaxy
population which we now believe were the dominant population in the early
Universe. Some of the galaxies detected this way would otherwise be completely un-detected even in deep HST photometry
(\citealt{Bacon15}, \citealt{Drake16}). Given a
statistical sample of objects, one can characterise the population by
examining the number of objects as a function of their luminosity -
the luminosity function, often fit with a Schechter function
parametrised by $\alpha$, $\phi^*$ and $L^*$ \citep{Schechter1976}. This gives us
characteristic values of the faint-end slope, number density, and
luminosity, respectively, and can be used to describe the nature of
the population at a particular redshift and to examine its evolution.

An efficient way to select large samples of star-forming galaxies is through
narrowband selection, whereby a narrow filter is used to select
objects displaying an excess in flux relative to the corresponding broad
band filter (e.g. thousands of H$\alpha$,  H$\beta$, [O{\sc{iii}}] and [O{\sc{ii}}]
emitters at $z < 2$ have been presented in \citealt{Sobral2013}, \citealt{Drake2013} and \citealt{Drake2015}).

Despite the success of
this technique, at high redshift where the Ly$\alpha$ line is
accessible to optical surveys the samples rarely probe far below $L^*$ and it is usually necessary
to make some assumption as to the value of the faint-end slope
$\alpha$, a key parameter for assessing the number of faint galaxies
available to power reionisation in the early Universe. (\citealt{Rhoads2000}, \citealt{Ouchi2003},
\citealt{Hu2004}, \citealt{Ouchi2008}, \citealt{Yamada2012}, \citealt{Matthee2015},
\citealt{Konno2015}, \citealt{Santos2016})

An alternative approach to LAE selection, was pioneered by
\cite{Martin08}, and subsequently exploited in \cite{Dressler11},
\cite{Henry12} and \cite{Dressler15}, combining
a narrowband filter with 100 long slits -- multislit narrowband
spectroscopy. With this technique the authors found evidence
for a very steep faint-end slope of the luminosity function, confirmed
by each subsequent study, although
they note that this is a sensitive function of the correction they
make for foreground galaxies. With this method, and taking fiducial
values for the escape of Ly$\alpha$, the escape of Lyman continuum
(LyC) and the clumping of the IGM they determined that LAEs probably produce a significant fraction of the ionising radiation required to maintain a transparent IGM at $z = 5.7$.

Some of the very deepest samples of LAEs to date come from blind long-slit spectroscopy, successfully discovering LAEs reaching
flux levels as low as  a few $\times 10^{-18} {\rm{erg\,s}}^{-1} {\rm{cm}}^{-2}$
for example \cite{Rauch2008}, although this required $92$ hours integration
with ESO-VLT FORS2, and the idenitifcation of many of the single line
emitters was again ambiguous. Reaching a similar flux limit, \cite{Cassata2011} combined targetted and
serendipitous spectroscopy from the VIMOS-VLT Deep Survey (VVDS) to
produce a sample of 217 LAEs across the redshift range {\mbox{$2.00 \le z
\le 6.62$}}. They split their sample into three large redshift bins,
and looked for signs of evolution in the observed luminosity function across the redshift range. In
agreement with \cite{vanBreukelen2005},
\cite{Shimasaku2006}, and \cite{Ouchi2008} they found no evidence of
evolution between the redshift bins within their errors. Due to the dynamic range
of the study, \cite{Cassata2011} fixed $\phi^*$ and $L^*$ in their
lowest two redshift bins and used the $1/$V$_{max}$ estimator to
measure $\alpha$, finding shallower values of
the faint-end slope than implied by the number counts of
\cite{Rauch2008} or \cite{Martin08}, \cite{Dressler11}, 2015, and
\cite{Henry12}. Similarly in the highest redshift bin they 
fixed the value of $\alpha$, fitting only
for $\phi^*$ and $L^*$, although interestingly they found the measured
value of $\alpha$ increased from their lowest redshift bin to the next. Deep spectroscopic studies are needed to evaluate the
faint end of the luminosity function, however each of the efforts to date suffer from
 irregular selection functions which are difficult to reproduce, and flux losses which are difficult to
quantify and correct for. 
\begin{figure*}
\begin{center}
\includegraphics[width=0.98\textwidth]{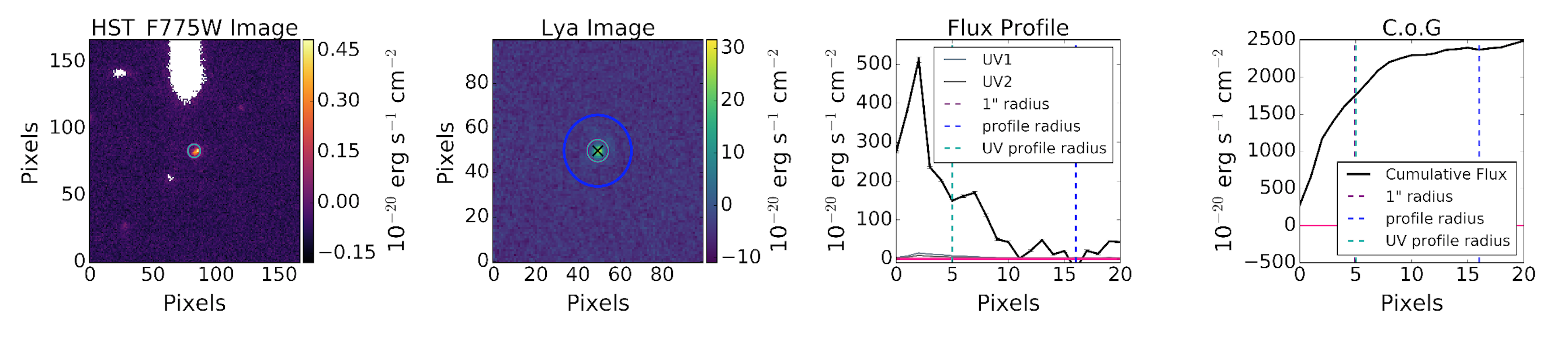} 
\caption[]{Example of flux estimation for object 149 in the
  {\emph{udf-10}} field. In the first panel we show the HST image
  corresponding to the wavelength of Ly$\alpha$, in
  the second panel we show the narrowband image extracted from the
  MUSE cube. In the third panel we show the flux profile of the galaxy
  determined according to the method described in Section \ref{sect:cog},
  and in the fourth panel we show the cumulative flux determined by
  summing the results in Section \ref{sect:cog}. The dashed vertical lines
  in the third and fourth panels show the $1 \arcsec$ radius, and the
  different radii encompassing the total flux according to a curve of
  growth analysis on either the HST or the MUSE images.}
\label{fig:flux}
\end{center}
\end{figure*}

The advent of the Multi Unit Spectroscopic Explorer (MUSE; \citealt{Bacon2010}), on the Very Large
Telescope (VLT), provides us with a way to blindly select large samples of
spectroscopically-confirmed LAEs in a homogeneous
manner (\citealt{Bacon15}, \citealt{Drake16}) as well as probing the
lensed LAE
populations behind galaxy clusters (\citealt{Bina2016}, \citealt{Smit16}). In addition to the ability of MUSE to perform as a `detection
machine' for star-forming galaxies, the deep datacubes allow us to
establish reliable total Ly$\alpha$ line fluxes by ensuring we capture the
total width of the Ly$\alpha$ line in wavelength, and the full extent
of each object on-sky. Indeed, Ly$\alpha$ emission has now increasingly
been found to be extended around galaxies, which has implications for
our interpretation of the luminosity function. \cite{Steidel11} first
proposed extended Ly$\alpha$ emission may be a generic property of
high redshift star-forming galaxies, which was confirmed by
\cite{Momose2014} who found scale lengths $\approx5 - 10$ kpc, but the emission was not detectable
around any individual galaxy (see also \citealt{Yuma2013} and Yuma et
al., submitted, for
individual detections of metal-line
blobs at lower $z$ from the sample of \citealt{Drake2013}). \cite{Wisotzki16} were the first to make detections of extended
Ly$\alpha$ halos around individual high-redshift galaxies, uncovering
$21$ halos amongst $26$ isolated LAEs presented in \cite{Bacon15}.\\

In this paper we build on the procedure developed in \citealt{Drake16} (hereafter D17a)
and upgrade our analysis in the ways outlined below. We push our
detection software {\sc{muselet}} to lower flux limits by tuning the
parameters. We incorporate a more sophisticated completeness
assessment than in D17a, by simulating extended Ly$\alpha$ emitters
and performing a fake source recovery experiment. We test and
correct for the effect of bias in our flux measurements of faint
sources, and finally we implement two different approaches to
assessing the Ly$\alpha$ luminosity function. The paper proceeds as
follows: in Section \ref{sect:data} we introduce our survey of the MUSE {\emph{Hubble Ultra
Deep Field; HUDF}} (\citealt{Bacon2017}; hereafter B17) and our catalogue construction from the parent data
set (presented in \citealt{Inami2017}; hereafter I17). In Section
\ref{sect:fluxes} we describe our approach to measuring accurate total
Ly$\alpha$ fluxes, and describe our method for constructing realistic
extended LAEs in Section \ref{sect:sim} to assess the possible bias
introduced in our flux measurements and to evaluate the completeness
of the sample. In Section \ref{sect:LFs} we present two alternative
approaches to assessing the Ly$\alpha$ luminosity function, first using the
$1/$V$_{max}$ estimator, and secondly a maximum likelihood approach to
determine the most likely Schechter parameters describing the
sample. We discuss in Section \ref{sect:discuss} the evolution of the
observed luminosity function, the contribution of LAEs to the overall
star formation rate density across the entire redshift range, and
finally the ability of LAEs to produce enough ionising radiation to
maintain an ionised IGM at redshift {\mbox{$5.00 \le z < 6.64$}}. 

Throughout the paper we assume a $\Lambda$CDM cosmology, {\mbox{$H_0 = 70.0$ kms$^{-1}$ Mpc$^{-1}$}},
{\mbox{$\Omega_{m} =0.3$}}, {\mbox{$\Omega_\Lambda = 0.7$}}. \\

\section{Data and catalogue construction}
\label{sect:data}

\subsection{Observations}
As part of the MUSE consortium Guaranteed Time Observations we
observed the Hubble Ultra Deep Field ({\emph{HUDF}}) for a total of 137 hours
in dark time with good seeing conditions (PI: R.Bacon). The observing
strategy consisted of a ten-hour integration across a $3\arcmin
\times 3\arcmin$ mosaic consisting of $9$ MUSE pointings, and overlaid on this, a
$30$ hour integration across a single MUSE pointing ($1\arcmin \times
1\arcmin$).Details on observing strategy and data reduction are given in Bacon et
al. 2017 (hereafter B17). MUSE delivers an instantaneous wavelength range of
{\mbox{$4750 - 9300\,\AA$}} with a mean spectral resolution of R $\approx 3000$, and
spatial resolution of 0.202\arcsec pix$^{-1}$. 

\begin{figure*}
\begin{center}
 \includegraphics[width=0.7\textwidth]{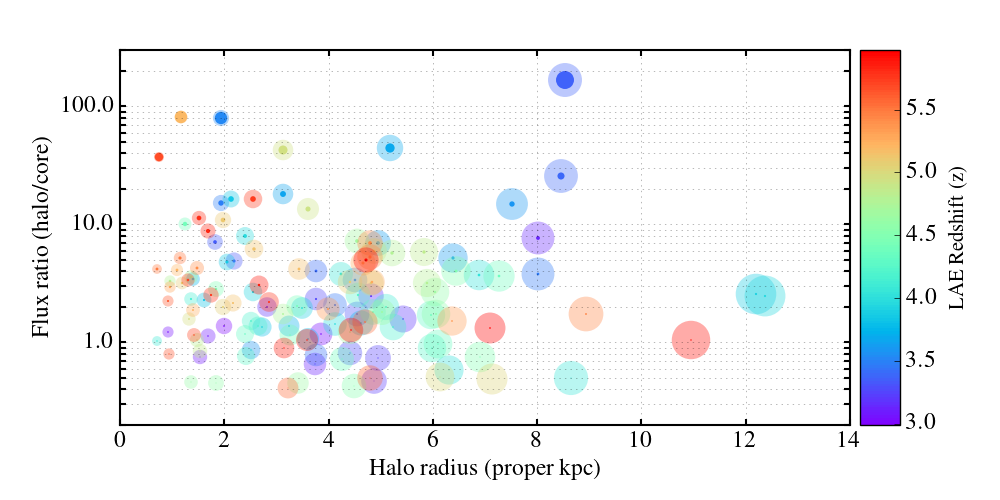} 
\caption[]{Halo properties used to simulate our realistic fake
  LAEs. The halos are entirely characterised by two quantities, the
  halo scale length in proper kpc, and the flux ratio between the halo
  and the core. Measurements are taken from those presented in L17 and
  W16. Each halo is depicted as an extended
disk of size proportional to the halo extent, overlaid with a compact
component of size inversely proportional to the flux ratio, this gives
an easy way to envisage the properties of the observed halos.}
\label{fig:halos}
 \end{center}
\end{figure*}

\subsection{Catalogue construction}
As we discussed at length in D17a, to assess the luminosity
function it is imperative to construct a sample of objects using a
simple set of selection criteria which can easily be reproduced when
assessing the completeness of the sample. Without fulfilling this
criterion it is impossible to quantify the sources missed during source detection and therefore impossible to reliably evaluate the luminosity function. For this reason we do not
rely solely on the official MUSE-consortium catalogue release (I17) - while the catalogue is
rich in data and deep, the methods employed to detect sources are
varied and heterogeneous, resulting in a selection function which is
impossible to reproduce. We instead choose to implement a single piece of detection software,
{\mbox{{\sc{muselet}}}}{\footnote{publicly available with MPDAF, see https://pypi.python.org/pypi/mpdaf for details}}, (J.Richard) and validate our detections through a
full 3D match to the deeper catalogue of I17. We note
that detection alogithms in survey data always require some
trade off to be made between sensitivity and the number of false
detections, and with a view to assessing the
luminosity function, the need for a well-understood selection function outweighs the need to detect
the faintest possible candidates (which are in principle ambiguous,
producing a less certain result than with a fully characterised
selection function).

We follow the procedure
outlined in D17a to go from a catalogue of {\mbox{{\sc{muselet}}}} emission-line
detections to a
catalogue of spectroscopically confirmed LAEs. The details are
outlined below, and further information can be found in D17a.

\subsubsection{Source detection}
\label{sect:source_det}
{\sc{muselet}} begins by processing the
entire MUSE datacube applying a
running median filter to produce continuum-subtracted narrowband
images at each wavelength plane. Each image is a line-weighted average
of $5$ wavelength planes ($6.25 \AA$ total width) with continuum
estimated and subtracted from two spectral medians on either side of
the narrowband region ($25 \AA$ in width). {\mbox{{\sc{muselet}}}} then runs the {\mbox{{\sc{SExtractor}}}} package
\citep{BertinArnouts1996} on
each narrowband image as it is created using the exposure map cube as
a weight map. 

Once the entire cube is processed, {\mbox{{\sc{muselet}}}} merges the detections from each narrowband image to create a
catalogue of emission lines. Lines which are co-incident on-sky within
$4$ pixels ($0.8 \arcsec$) are merged into single sources, and an
input catalogue of rest-frame emission-line wavelengths and flux ratios is used to determine a best redshift
for sources with multiple lines, the remainder of sources displaying a
single emission line are flagged as
candidate LAEs. Thanks to the wavelength coverage of MUSE, we
anticipate the detection of multiple lines for sources exhibiting
H$\alpha$, H$\beta$ or [O{\sc{iii}}] emission meaning that only the
[O{\sc{ii}}] doublet is a potential contaminant of the single-line
sample.

\subsubsection{Final catalogue}
Each of our {\mbox{{\sc{muselet}}}} detections is validated
through a 3D match to I17, requiring sources to be coincident on-sky
($\Delta$\,RA, $\Delta$\,DEC $< 1.0\, \arcsec$) and in observed wavelength
($\Delta\, \lambda < 6.25 \AA$).

We investigated the setup of
both {\mbox{{\sc{SExtractor}}} and {\mbox{{\sc{muselet}}}} parameters that would
optimise the ratio of matches to the total number of {\mbox{{\sc{muselet}}}}
detections. The results of these experiments led to our lowering the  {\mbox{{\sc{muselet}}}} clean threshold
to $0.4$ - meaning that only parts of the cube with fewer than $40\%$ of
the total number of exposures were rejected by the
software. Additionally, the {\mbox{{\sc SExtractor}}} parameters {\mbox{{\sc detect minarea}}}, and
{\mbox{{\sc detect thresh}}} (minimum number of contiguous pixels above the
threshold and {\mbox{detection-$\sigma$}} respectively) were each
lowered to $= 2.0$ from our previous stricter requirements of $3.0$
and $2.5$ in D17a. Naturally, the cost of lowering our detection
    thresholds is to increase the number of false detections from
    {\mbox{{\sc{muselet}}}} (which were negligible in the pilot study), and as such our completeness estimates
    here could be slightly overestimated.

Our match to I17 confirmed that $123$ and $481$ single line sources were LAEs in
the {\emph{udf-10}} and {\emph{mosaic}} fields respectively. In Figure \ref{fig:nocounts} we
show the parent sample from I17 in blue, overlaid with the
{\sc{muselet}}-detected sample depicted by a black hatched
histogram. In the two left-hand panels we show LAEs as a function of redshift demonstrating a flat distribution of
objects across the entire redshift range, and no sytematic bias in the
way we select our sample. In the two right-hand panels we show LAEs as
a function of the Ly$\alpha$ flux estimates presented in I17. With our re-tuning of the
{\sc{muselet}} software we now recover LAEs as faint as
a few $\times 10^{-18} {\rm{erg\,s}}^{-1} {\rm{cm}}^{-2}$. 

\begin{figure*}
\begin{center}
 \includegraphics[width=0.98\textwidth]{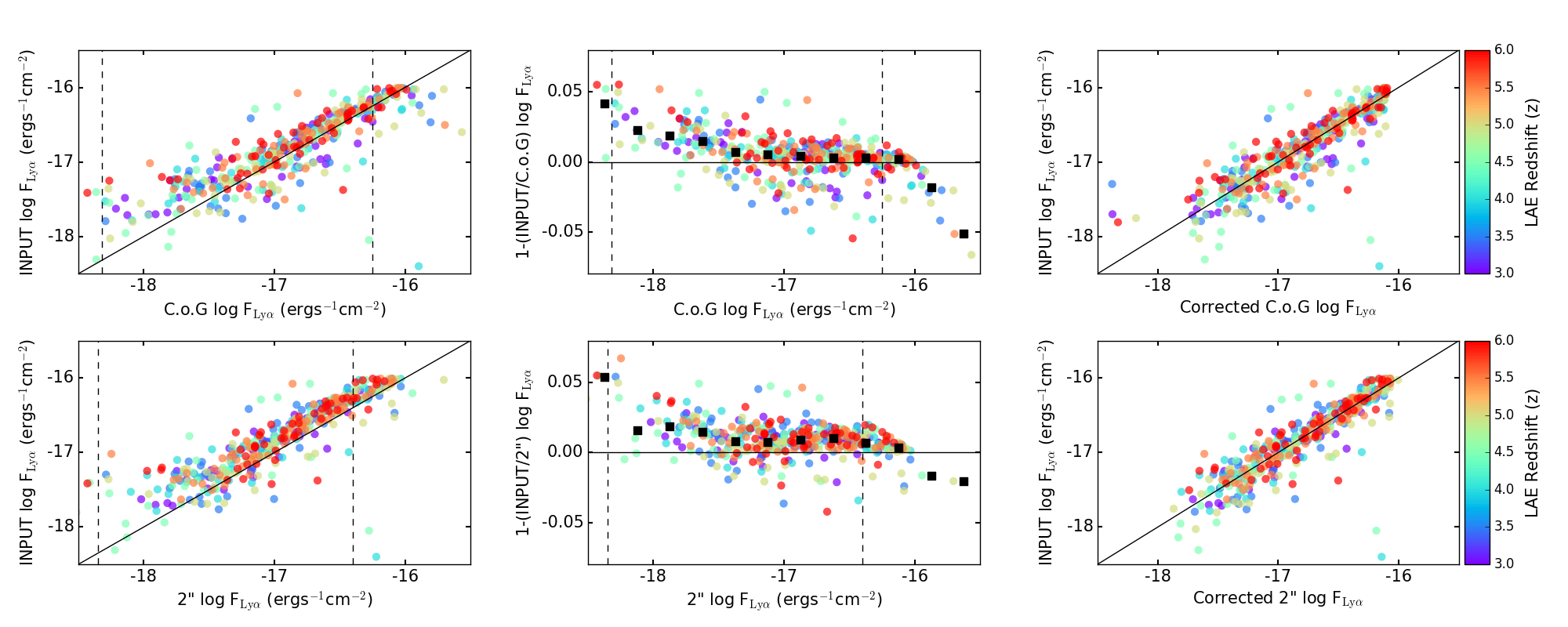} 
\caption[]{Bias in flux estimation for C.o.G (upper row) and 2''
  aperture (lower row) measurements in the {\emph{UDF-10}} field. In the first column of panels we
  show a comparison between the input total flux on the ordinate and
  the recovered flux on the abscissa. In the central column of panels
we show the difference between input and recovered flux on the
ordinate as a function of recovered flux on the
abscissa, where the black squares indicate the median value of the
offset, which increases rapidly towards lower fluxes. In each of the first two columns of panels we depict the
minimum and maximum fluxes of objects detected in the MUSELET
catalogue with dashed lines. In the
final column of panels we show values of measured flux
corrected for the median offset using measurements from the central columns.}
\label{fig:bias}
\end{center}
\end{figure*}

\section{Flux measurements}
\label{sect:fluxes}
The accurate measurement of Ly$\alpha$ fluxes has proved to be
non-trivial. Furthermore, the definition of the Ly$\alpha$ flux itself is
changing now that we are working in the regime where LAEs
are seen to be extended objects often with diffuse Ly$\alpha$-emitting
halos. Here we work mainly with our best estimates of the
total Ly$\alpha$ flux for each object, that is, including
extended emission in the halos of galaxies.

In section $3$ of D17a, we discussed the most accurate way to determine total Ly$\alpha$
fluxes and argued that a curve-of-growth approach provided the most
accurate estimates. Here, we again investigate the curve-of-growth
technique, but before developing a more advanced analysis, we consider the
possible bias that might be inherent to this method in our ability
to fully recover flux according to the true total flux. The approach
developed to correct for this bias is described in Section
\ref{sect:bias}. 

This work upgrades the preliminary analysis presented in D17a to make use of the MUSE-{\emph{HUDF}}
data-release source objects. For each source found by
{\sc{muselet}} with a match in the catalogue of I17 we take the source objects provided in the data release, and
measure the FWHM of the Ly$\alpha$ line on the 1D spectrum. We then
add two larger cutouts of $20 \arcsec$ on a side to each source object from the
full cube -- a narrowband and a continuum image. The narrowband image, centred on the wavelength of the
detection, is of width $4 \times$ the FWHM of the line, and the
continuum image is $200 \AA$ wide, offset by $150 \AA$ from the
peak of the Ly$\alpha$ detection. By subtracting the broadband from the
narrowband image we construct a ``Ly$\alpha$ image''
(continuum-subtracted narrowband image) and it is on this image that
we perform all photometry. 

\subsubsection{Curve of growth}
\label{sect:cog}
We use the python package {\sc{photutils}} to prepare the Ly$\alpha$
image by performing a local background
subtraction, and masking neighbouring objects in
the Ly$\alpha$ image. Then taking the {\sc{muselet}}
detection coordinates to be the centre of each object, we place consecutive annuli of increasing radius on the
object, taking the average flux in each ring as we go, multiplied by
the full area of the annulus. When the average value
in a ring reaches or dips below the local background, we sum the flux
out to this radius as the total Ly$\alpha$ flux. 

\subsubsection{Two arcsecond apertures}
We prepare the image in the same way as for the curve-of-growth analysis, and
again take the {\sc{muselet}} coordinates as the centre of the Ly$\alpha$
emission. Working with the same set of consecutive annuli we simply
sum the flux for each object when the diameter of the annulus reaches
$2 \arcsec$. We note that this produces an ever so slightly different
result to placing a $2 \arcsec$ aperture directly on the image.

\section{Simulating realistic extended LAEs}
\label{sect:sim}
In D17a we based our fake source recovery experiments on
point-source line-emitters using the measured line profiles
from the galaxies presented in our study of the Hubble Deep Field
South \citep{Bacon15}. While the estimates provided a handle on the
completeness of the study, we noted that the reality of extended
Ly$\alpha$ emission might make some significant impact on the recovery
fraction of LAEs (see Herenz et al). Additionally, our completeness estimates are based on the
input Ly$\alpha$ flux, and so it is prudent to understand the
relationship between measured fluxes and the most likely intrinsic
flux. To address
both the issue of completeness of extended LAEs and the question of
some bias in the recovery of total Ly$\alpha$ flux we designed a fake source recovery experiment using
``realistic'' fake LAEs. We model extended Ly$\alpha$ surface brightness profiles with no
continuum emission, making use of the detailed
measurements of \citealt{Leclercq2017} (hereafter L17) performed on all Ly$\alpha$ halos
detected in the MUSE {\emph{HUDF}} observations, and \cite{Wisotzki16} on those
in the {\emph{HDFS}}. Both L17 and \cite{Wisotzki16} follow a
similar procedure to decompose the LAE light
profiles, invoking a ``continuum-like'' core component, and a diffuse,
extended halo.

We approximate the central continuum-like component as a point source,
and combine this with an exponentially declining profile to represent
the extended halo. The emitters can then be entirely characterised by two parameters; the halo
scale length in proper kpc, and the flux ratio between the halo and the
core components. Figure \ref{fig:halos} shows the distribution of halo
parameters used in the simulation. The extent of the halo in proper
kpc is given on the abscissa, and the flux ratio between the extended
halo component and the compact continuum-like component is given on
the ordinate, with colours indicating the redshift of the halo
observed by \cite{Wisotzki16} or L16. We depict each halo as an extended
disk of size proportional to the halo extent, overlaid with a compact
component of size inversely proportional to the flux ratio, this gives
an easy way to envisage the properties of the observed halos. For each of our
experiments, described below, we draw halo parameters from the
measured sample in a large redshift bin ($\Delta z \approx 1$) centred
on the input redshift of the simulated halo. 

\begin{figure*}
\begin{center}
\includegraphics[width=0.98\textwidth]{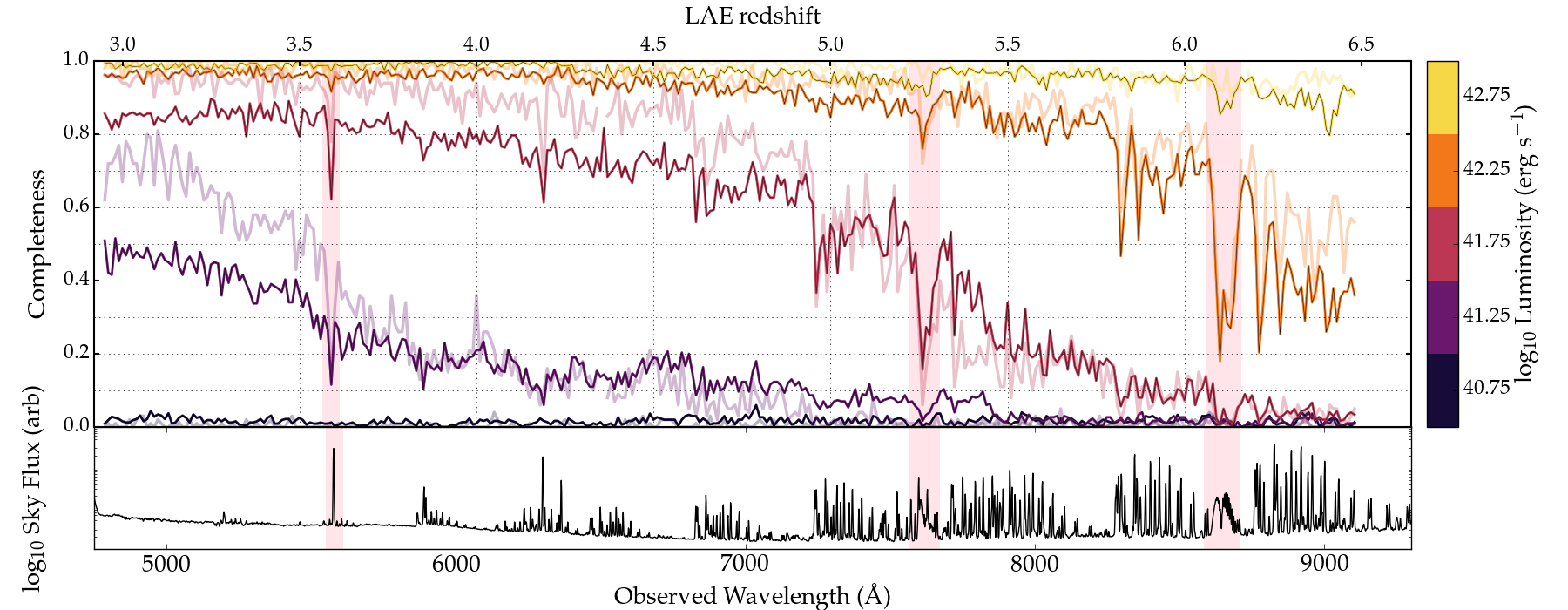} 
 \includegraphics[width=0.98\textwidth]{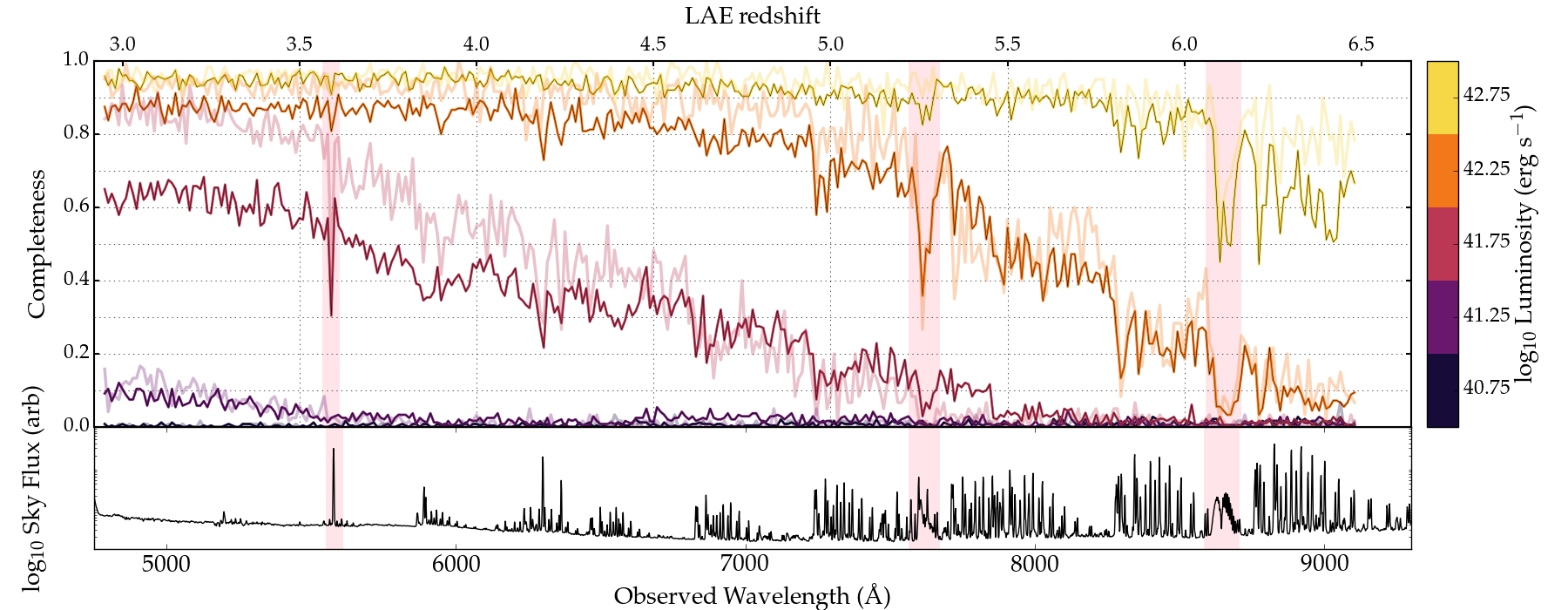} 
\caption[]{The recovery fraction of LAEs with our
       detection software as a function of observed wavelength, and
       LAE-redshift denoted on the top axis for the {\emph{udf-10}} (top),
       and the {\emph{mosaic}} (bottom) fields. Colours represent the
       input luminosity of the fake LAEs, dark lines reinforced in
       black show the recovery of extended objects, and pale lines
       show point sources of the same total luminosity. In the
     lower panel of each plot the night sky is shown, and areas where sky lines
     most severely affect our recovery are highlighted in pink.}
\label{fig:compl}
\end{center}
\end{figure*}
\subsection{Flux recovery of simulated emitters}
\label{sect:bias}
In \cite{Drake16} we discussed the difference in the apparent
luminosity function when using different approaches to estimate total
Ly$\alpha$ flux. We concluded that using a curve-of-growth analysis
provided the most accurate measure of F$_{\rm{Ly\alpha}}$ although
noted that this approach introduced the possibility of a bias in the
fraction of the flux recovered according to true total flux. \\

Here, we inserted fake sources with a wide range of input fluxes and randomly
drawn halo parameters at a series of discrete redshifts. For those
objects that were recovered by our detection software we could then
apply the same methods of flux estimation that we employed for the
sample of real objects to uncover any systematic bias in the way we
estimate total Ly$\alpha$ fluxes. The results of this experiment are
presented in Figure \ref{fig:bias} for the {\emph{udf-10}} field.

Interestingly, the curve-of-growth recovers the total input flux
remarkably well at bright fluxes, but has a
huge scatter at lower fluxes rendering it completely unreliable,
although not systematically wrong. Secondly, the 2$\arcsec$
measurements seem to work fairly well at lower
fluxes, but diverge systematically ar higher flux levels (as we
discussed in D17a).

For each of these two approaches to flux recovery, we calculate the
median offset of the recovered fluxes from the input fluxes, and
interpolate the values in order to make statistical correction as a
function of recovered flux to the
measured values. In the final column of panels in Figure
\ref{fig:bias} we show the corrected values for first the
curve-of-growth, and then the $2\arcsec$ apertures for measurements on
the {\emph{udf-10}} field. It can be seen in these
plots that while both estimates are now centred on an exact
correlation between input and recovered flux, the scatter in the
$2\arcsec$ measurements is much lower than that in the corrected
curve-of-growth values. For this reason it is the corrected
$2\arcsec$ aperture flux values which we propagate to the
luminosity functions. We find a typical offset of $0.02$ in
      log F (erg s$^{-1}$ cm$^{-2}$)
      with an average r.m.s of $0.008$. 

\begin{figure*}
\begin{center}
 \includegraphics[width=0.48\textwidth]{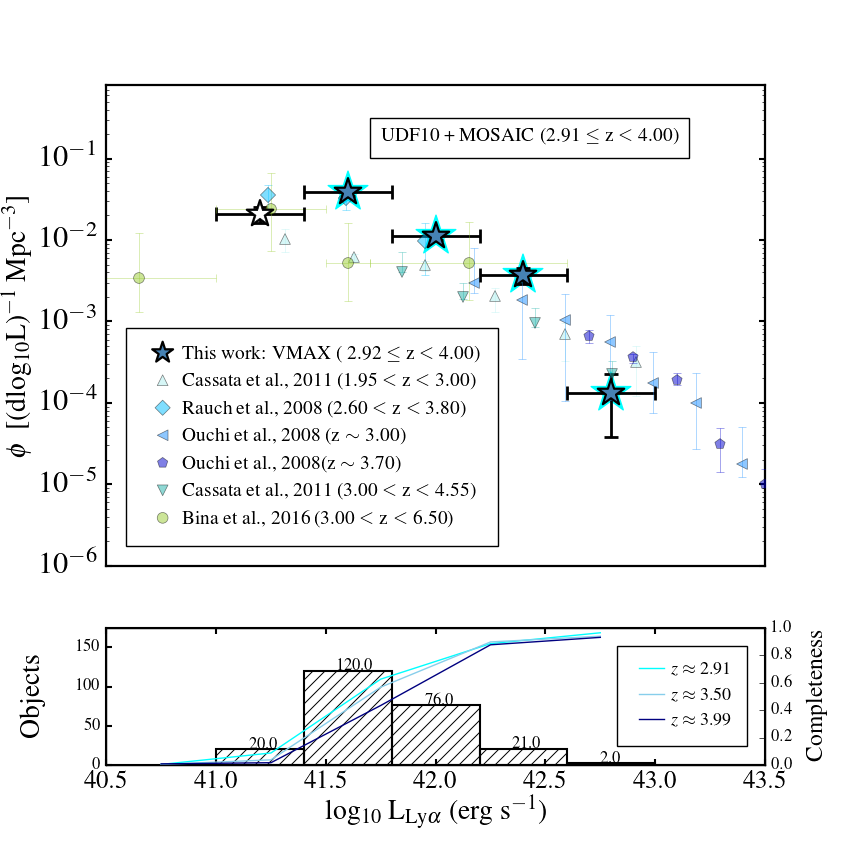}
 \includegraphics[width=0.48\textwidth]{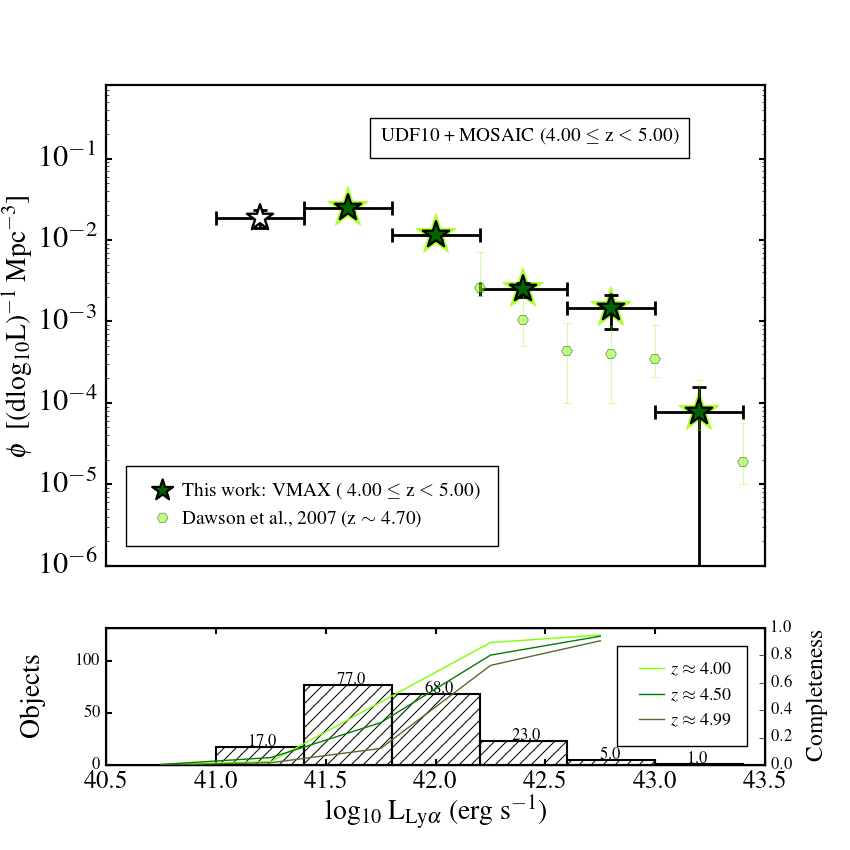}
 \includegraphics[width=0.48\textwidth]{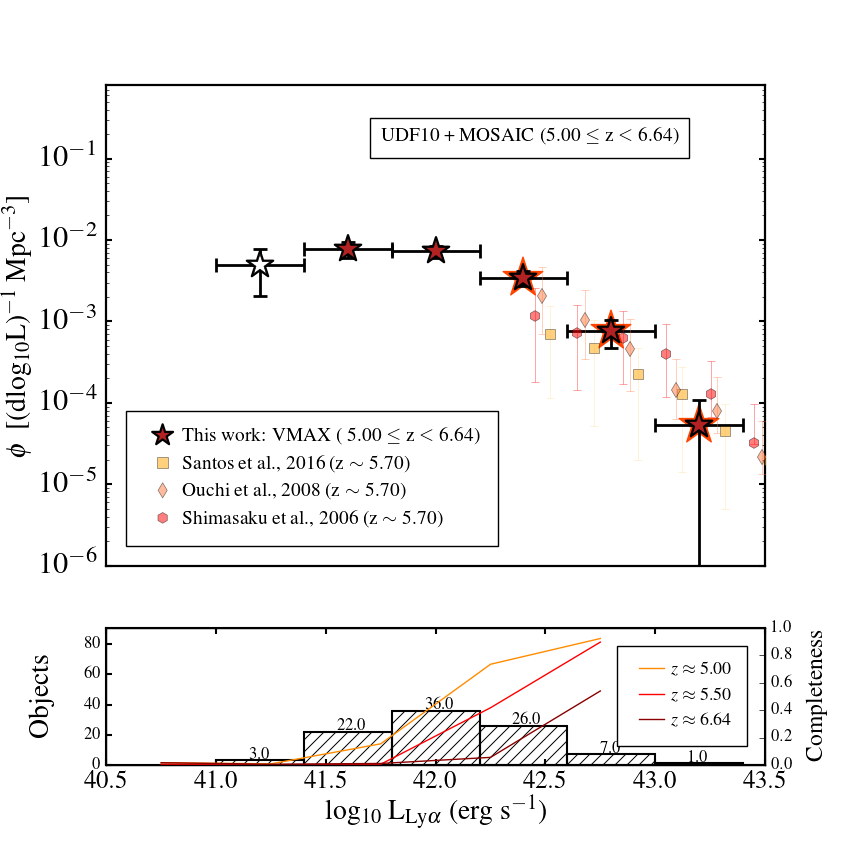}
 \includegraphics[width=0.48\textwidth]{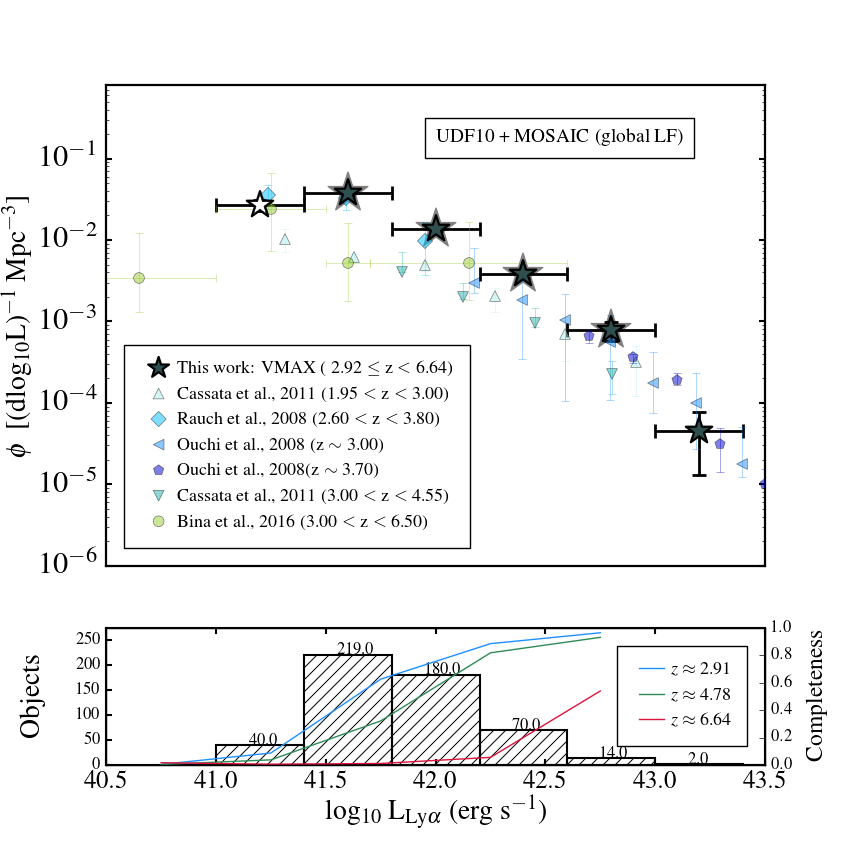}
\caption[]{Number densities resulting from the $1/$V$_{max}$
  estimator. Top left: \mbox{($2.91 \le z < 4.00$) bin; blue}, top right \mbox{($4.00
\le z < 5.00$); bin green}, bottom left \mbox{($5.00
\le z < 6.64$); bin red}, bottom right all LAEs \mbox{($2.91
\le z < 6.64$)}. In each panel we show number densities in bins of
0.4 dex, together with literature results at similar redshifts from
narrowband or long-slit surveys. In the lower part of each panel we
show the histogram of objects in the redshift bin overlaid with the
completeness estimate for extended emitters at the lower, middle and
highest redshift in each bin. In
each panel we flag incomplete bins with a transparent
datapoint. Errorbars represent the $1\sigma$ Poissonian
uncertainty, we note that often the ends of the bars are hidden behind
the data point itself.}
\label{fig:Vmax_LFs_zbins}
\end{center}
\end{figure*}

\subsection{Fake source recovery}
\label{fake_recovery}
We follow the procedure described in D17a, working systematically
through the cube adding fake emitters in redshift intervals of $\Delta
z = 0.01$. This time we use two different setups designed to
facilitate two different approaches to estimating the luminosity
function - the first using 5 luminosity bins, and
the second using flux intervals of $\Delta f = 0.05$ (erg s$^{-1}$ cm$^{-2}$). The
incorporation of the completeness estimates into the luminosity
functions is described in Section \ref{sect:LFs}. For each fake LAE
inserted into the cube, observed pairs of values of
scale length and flux ratio are drawn from the measurements presented
in \cite{Wisotzki16} and L16. For each
redshift-flux and redshift-luminosity combination we run our detection software {\sc{muselet}} using
exactly the same setup as described in Section \ref{sect:source_det}, and
record the recovery fraction of fake extended emitters. \\

\section{Luminosity functions}
\label{sect:LFs}
Here we implement two different estimators to assess the luminosity
function, each with their own strengths and weaknesses. With a view to
estimating the number density of objects in bins of luminosity, the
$1/$V$_{\rm{max}}$ estimator provides
a simple way to visualise the values and makes no prior assumption as
to the shape of the function. We discuss
the limitations of this approach in our pilot study of the HDFS; D17a. In terms of parameterising the luminosity function, fits to binned
data are to be interpreted with caution, and an alternative approach is
preferred. Here, we use the maximum likelihood estimator following the formalism described in
\citealt{Marshall83} (and applied in \citealt{Drake2013} and
\citealt{Drake2015} to narrowband samples).

One advantage of the MUSE {\emph{mosaic}} of the HUDF is that the $3 \times
3$ square arcminute field in combination with
the $10$-hour integration time, provides the ideal volume to
capitalise on the trade-off between minimising cosmic
variance and probing the bulk of the LAE population \citep{Garel2016}. This
allows us to draw more solid conclusions than those from our $1 \times
1$ pilot study of the HDFS field \citep{Drake16}.

\subsection{1/Vmax estimator}
\label{sect:vmax}
We assess the luminosity
function in $3$ broad redshift bins {\mbox{($2.91 \le z < 4.00$)}}, {\mbox{ ($4.00 \le z \le
4.99$)}} and {\mbox{ ($5.00 \le z < 6.64$)}} in addition to the `global' luminosity
function {\mbox{ ($2.91 \le z \le  6.64$)}} for LAEs in the combined
{\emph{UDF-10}} plus {\emph{mosaic}} field through use of the
$1/V_{{\rm{max}}}$ estimator. The results, discussed further below, are
presented in Table {\ref{tab:vmax}} and
Figure \ref{fig:Vmax_LFs_zbins}. 

\begin{figure*}
\begin{center}
 \includegraphics[width=0.98\textwidth]{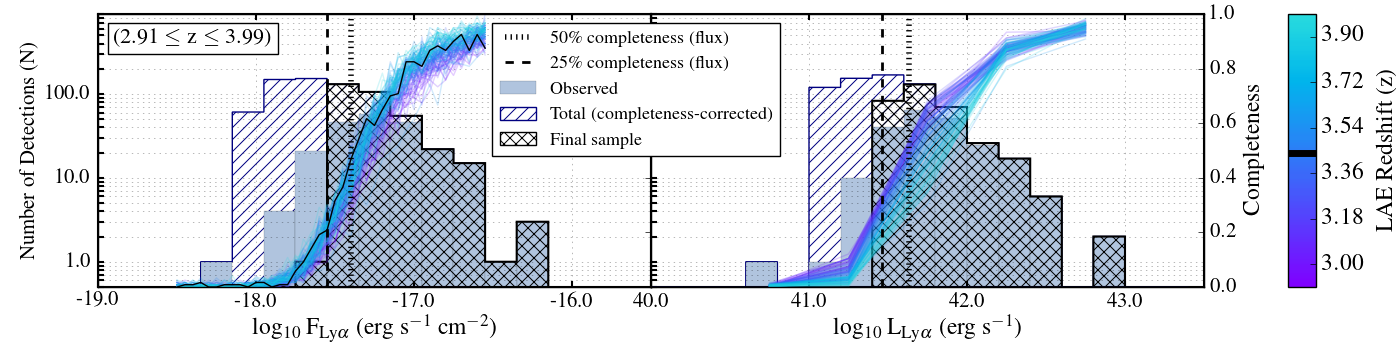} 
 \includegraphics[width=0.98\textwidth]{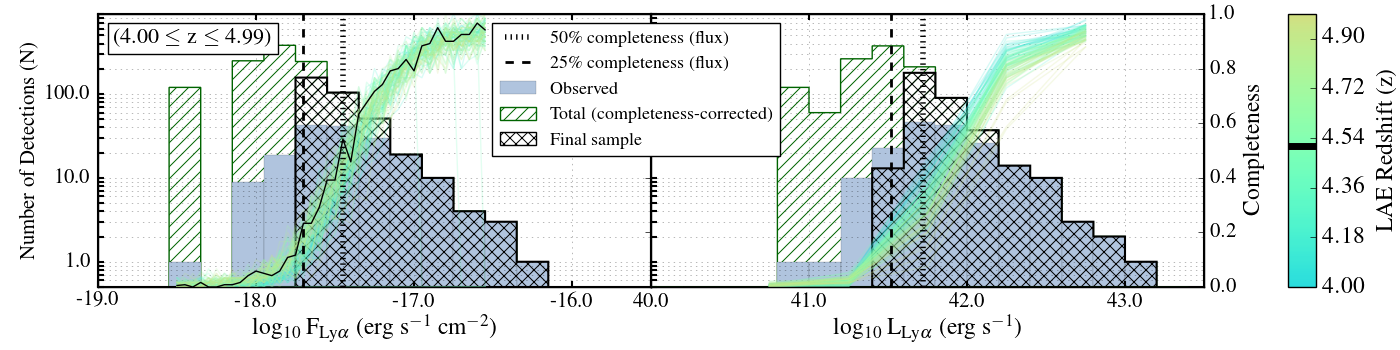} 
 \includegraphics[width=0.98\textwidth]{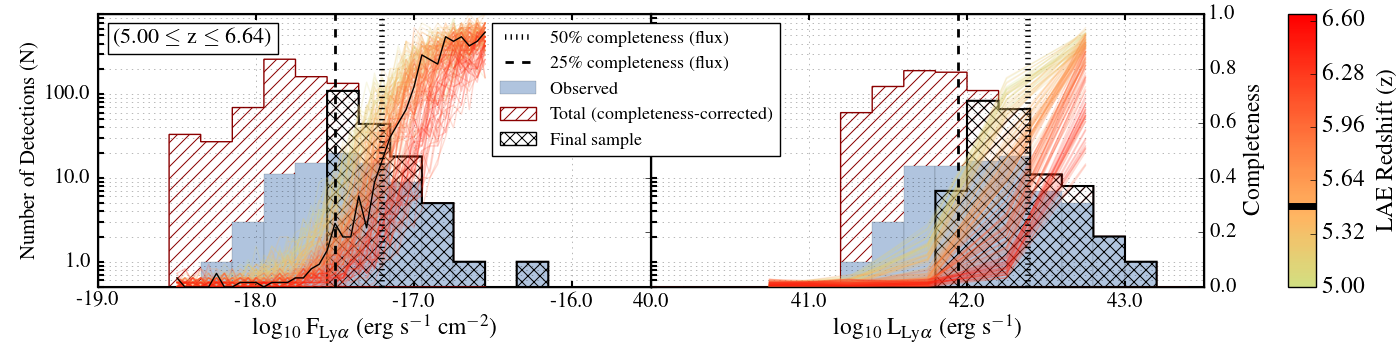} 
\caption[]{Flux and
log-luminosity distributions of
objects from the {\emph{mosaic}} in three broad redshift
bins at  {\mbox{($2.91 \le z < 4.00$)}}, {\mbox{ ($4.00 \le z \le
4.99$)}} and {\mbox{ ($5.00 \le z < 6.64$)}}. In each
panel we show the total distribution of objects (including fakes
created and added to the sample through the process described in
Section \ref{sect:ML_LF})
in a coloured hatched histogram. We overlay the distribution
of observed objects in filled blue bars. The final samples, curtailed at the $25\%$ completeness limit in flux ($\Delta\,f =
0.05$) for the median redshift of objects in the redshift bin is
overplotted in a bold cross-hatched black histogram. Overlaid on each panel are the completeness curves as a function of flux (or log luminosity) at
each redshift ($\Delta\,z = 0.01$) falling within the bin. Each
redshift is given by a different coloured line according to the
colour-map shown in the colour bar, and the curve
at the median redshift of the bin is emphasized in black. The median
redshift of the bin is also given by a black line on the colour bar.}
\label{fig:sample2}
\end{center}
\end{figure*}

\subsubsection{Completeness correction}
To implement the $1/$Vmax estimator, it is necessary to evaluate the
completeness of the sample for a given luminosity as a function of redshift. In Figure \ref{fig:compl} we show the recovery fraction of LAEs
with {\sc muselet} as a function of observed wavelength at $5$ values of log
luminosity, giving the corresponding LAE-redshift on the top axis. The night sky spectrum
from MUSE is shown in the lower panel, and colour-coding
of the lines represents the in-put luminosity of the sources ranging
between {\mbox{$41.0 <$ log $L$ (erg s$^{-1}) < 43.0$}} at each wavelength of the cube in intervals of $\Delta
\lambda = 12 \AA$. In the upper panel we show the recovery fraction from the deep
$1\arcmin \times 1\arcmin$ {\emph{udf-10}} pointing inserting $20$ LAEs at a time
in a $z$-$L$ bin.

The effects of night sky emission are most evident at
luminosities up to log $L$$\approx 42.5$. The prominent [O{\sc{i}}] airglow line at
$5577\AA$ however impacts recovery even at the brightest luminosities in our
simulation. The broader absorption features at $7600\AA$ and
$8600\AA$ also make a strong impact on detection efficiency across
the full range of luminosities. Importantly, the difference between the recovery fractions of
point-like and extended emitters is evident. For each coloured line of
constant luminosity, we show two different recovery fractions; the extended emitter recovery fraction, and the point-source recovery
fraction. It is obvious that for a
given total luminosity the point-like emitters are recovered more
readily than the extended objects meaning that our previous
recovery experiements will have overestimated the completeness of the
sample.

\begin{figure*}
\begin{center}
 \includegraphics[width=0.7\textwidth]{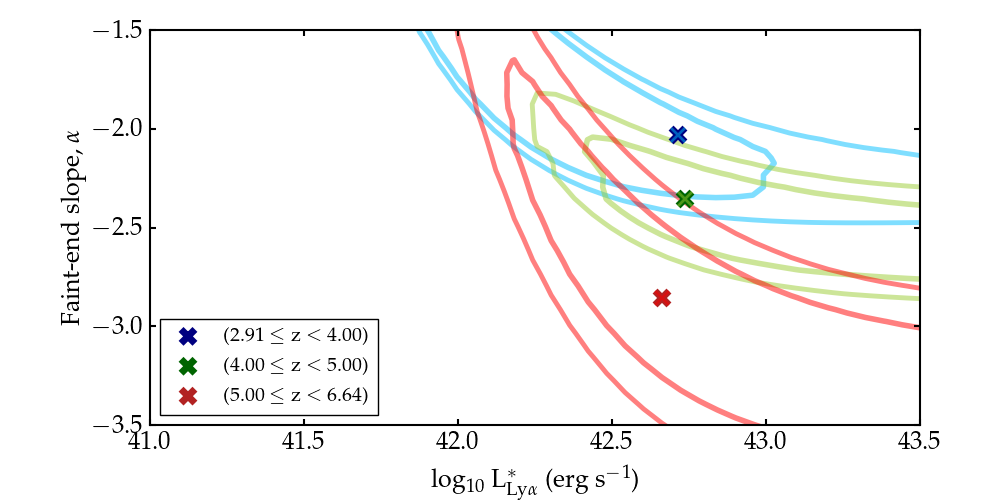} 
 \includegraphics[width=0.7\textwidth]{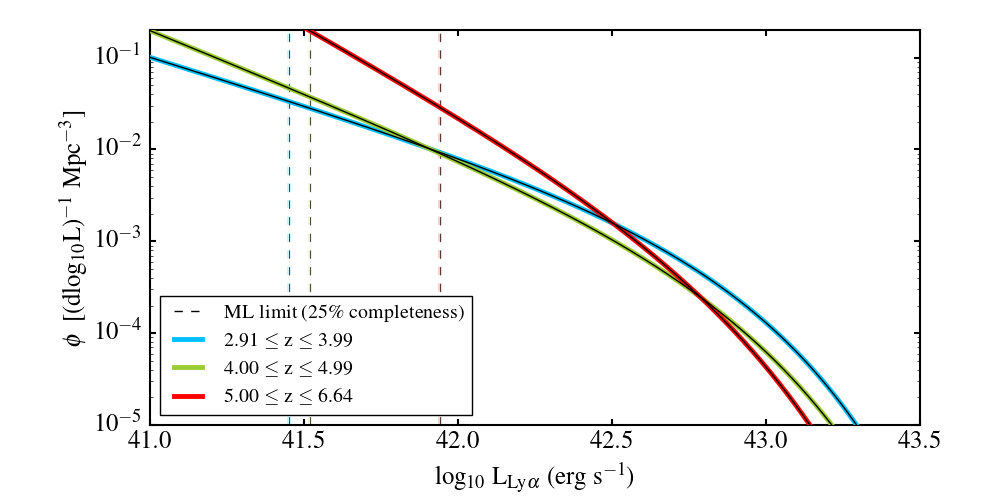} 
\caption[]{Maximum likelihood Schechter luminosity functions for three redshift
  bins {\mbox{($2.91
\le z < 4.00$); blue}}, {\mbox{($4.00
\le z < 5.00$); green}}, {\mbox{($5.00
\le z < 6.64$); red}}. In the upper panel
we show 68\% and 95\% joint confidence regions which
correspond to $\Delta$ S $= 2.30$ and $6.18$ for two free fit parameters (L$^*$ and $\alpha$). In the lower panel we show the maximum likelihood Schechter functions as solid lines.}
\label{fig:Banana}
\end{center}
\end{figure*}

\subsubsection{1/Vmax formalism}
For each LAE, {\emph{i}}, in the catalogue, the
redshift, $z_{\,i}$, is determined according to {{$z_{\,i} = {\lambda_{\,i}}/({1215.67} - 1.0)$}}, where $\lambda_{\,i}$ is the observed wavelength of
Ly$\alpha$ according to the peak of the emission detected by {\sc
  muselet}. The luminosity $L_{i}$ is then computed according to
{{$L_i = f_{i} 4 \pi D_{L}^{2} (z_{i})$}}, where $f_{i}$ is the
corrected Ly$\alpha$ flux measured in a $2 \arcsec$ aperture, $D_{L}$ is the luminosity distance, and $z_{i}$ is the
Ly$\alpha$ redshift. The maximum co-moving volume within which this object could
be observed, $V_{{\rm{max}}}$($L_{\:i}$, $z$), is then computed by:

\begin{equation}
{V_{{\rm{max}}}(L_{i}, z) = \int_{z1}^{z2} \frac{{\rm{d}}V}{{\rm{d}}z}\, {\rm{C}}(L_{i},z)\, {\rm{d}}z,}
\label{eq:Vmax}
\end{equation}

\noindent where $z_1$ and $z_2$, the minimum and maximum redshifts
of the bin respectively, d$V$ is the co-moving volume element corresponding to redshift interval
d$z = 0.01$, and C($L_{i}$, $z$) is the completeness curve for an extended
object of total luminosity $L_i$, across all redshifts $z_i$.

The number density of objects per luminosity bin, {\rm{$\phi$}}, is then calculated according to:

\begin{equation}
{\phi [({\rm{d log_{10}}} L)^{-1} {\rm{Mpc}}^{-3}] = \sum_i \frac{1}{V_{\rm{max}}(L_{i}, z_{i})} / {\rm{bin size}}.}
\label{eq:phi}
\end{equation}

\subsubsection{1/Vmax comparison to literature}
With an improved estimate of completeness
from the realistic extended emitters we see in Figure \ref{fig:Vmax_LFs_zbins} that the LF is steep down
to log luminosities of $L$$< 41.5$ erg s$^{-1}$, and sits
increasingly higher than literature results towards fainter
luminosities. This is entirely expected due to our improved
completeness correction following the analysis in D17a, and consistent
with the scenario in which the ability of MUSE to capture extended
emission results in a luminosity function showing number densities
systematically above previous literature results by a factor of
$2$-$3$. In each panel of Figure \ref{fig:Vmax_LFs_zbins} the redshift
 range is given in the upper right-hand corner, number densities from
 this work are depicted, plotted together with
 literature data across a similar redshift range identified in the
 key. Each data point from MUSE is shown with a Poissonian error on the point. In the lower part of each panel we show the histogram of
 objects' luminosities in the redshift bin, and overplot the the
 completeness as a function of luminosity at the lowest, central and
 highest redshifts contained in the luminosity function. This is
 intended to allow the reader to interpret each luminosity function
 with the appropriate level of caution - for instance in the highest
 redshift bin more than half the bins of luminosity consist of objects
 where a large completeness correction will have been used on the
 majority of objects, and hence there is a large associated
 uncertainty.

In the {\mbox{($2.91 \le z < 4.00$)}} bin, our data alleviate the
discrepancy between the two leading studies at redshift $\approx 3$
from VVDS \citep{Cassata2011} and \cite{Rauch2008}. Our
data points sit almost exactly on top of those from \cite{Rauch2008}
confirming that the majority of single line emitters detected in their
$90$-hour integration were LAEs. In the  {\mbox{ ($4.00 \le z < 5.00$)}} bin our data are over $1$
dex deeper than the previous study at this redshift \citep{Dawson2007},
we are in agreement with their number densities within our error bars
at all overlapping luminosities, and our data show a continued steep
slope down to $L$$< 41.5$ erg s$^{-1}$. In our highest redshift bin, {\mbox{ ($5.00 \le z < 6.64$)}}, our
data are a full $1.5$ dex deeper than previous studies. The data turn
over in the bins below $L$$< 42$ erg s$^{-1}$ but errors from the
completeness correction to objects in these bins is large since
values of completeness are well below $50\%$ for all luminosities in
the bins in this redshift range. Finally we show the `global' luminosity function across the redshift
range {\mbox{ ($2.91 \le z \le  6.64$)}} in the final panel together
with literature studies that bracket the same redshift range, and the
two narrowband studies from {\cite{Ouchi2003}} and {\cite{Ouchi2008}}
which represent the reference samples for high-redshift LAE studies. 

\begin{table*}
\caption{Maximum likelihood Schechter luminosity functions for LAEs in the
  {\emph{mosaic}} field. Marginal 68\% confidence intervals
        on single parameters are taken from the extremes of the
        $\Delta\, S = 1$ contours. The 68\% confidence intervals on the luminosity density and SFRD
    however depend on the joint confidence interval the
    two free parameters L$^*$ and $\alpha$ ($\Delta\, S = 2.30$
    contour as in Figure \ref{fig:Banana}) - details in Section
    \ref{sect:error_anal}. We note that as
      our sample are almost entirely below L$^*$, the value of L$^*$
      itself is only loosely constrained by our data, and hence we only
      find a single bound of the
68\% confidence intervals for the Schechter parameters in two redshift bins. Thankfully this is
not a problem for the luminosity
density and SFRD, as the extreme values are reached in a perpendicular
direction to the length of the ellipses.}
\label{tab:ML_sch_params} 
\centering            
\renewcommand{\arraystretch}{1.4}                   
\begin{tabular}{c c c c c c c c c }   
\hline \hline 
z & Volume & Real objects$^{\dagger}$ &Total$^{\dagger}$ & log$_{10}$$\phi^*$ & log$_{10}$$L^*$  & $\alpha$ &  log $\rho_{Ly\alpha}$$^{\dagger \dagger}$ &  SFRD$^{\dagger \dagger}$\\ 
  & $10^{4}$ Mpc$^{-3}$ &  & & (Mpc$^{-3}$) & (erg s$^{-1}$)  & & (erg s$^{-1}$Mpc$^{-3}$) & (M$_{\odot}$  yr$^{-1}$Mpc$^{-3}$) \\ 
\hline \hline 
\hline 
{\bf{(   2.92$\le z \le $   3.99)}} &   3.10  & 193 &     328 &   -3.10$^{+1.37 }_{-0.45}$
&   42.72$^{+0.23}_{-0.97}$ &   -2.03$^{+ 1.42}_{-0.07}$ &   40.154$^{+   0.346}_{-   0.138}$ &   0.014$^{+   0.017}_{-   0.004}$ \\ 
\hline 
{\bf{(   4.00$\le z \le $   4.99)}} &   2.57 &  144 &     346 &
-3.42$^{+0.51}_{-\infty}$ &   42.74$^{+ \infty}_{-0.19}$ & -2.36$^{+ 0.17}_{-\infty}$  &   40.203$^{+   0.397}_{-0.002}$ &   0.015$^{+   0.023}_{ -0.000}$ \\ 

{\bf{(   5.00$\le z \le $   6.64)}} &  3.64 &    50 &     176 &
-3.16$^{+0.99}_{-\infty}$ &   42.66$^{+\infty}_{-0.34}$ &   -2.86$^{+0.76}_{-\infty}$ &   40.939$^{+   0.591}_{-   0.727}$ &   0.083$^{+   0.240}_{-   0.067}$ \\ 

\hline 
\multicolumn{8}{l}{$\dagger > 25\%$ completeness in flux at the median
  redshift of the luminosity function.}\\
\multicolumn{8}{l}{$\dagger \dagger$ Integrated to log$_{10}$ L erg s$^{-1}$ $=41.0$}\\

\end{tabular}
\end{table*}

\subsection{Maximum likelihood estimator}
\label{sect:ML_LF}
With a view to parameterising the luminosity function we apply the
maximum likelihood estimator. Bringing together our bias-corrected flux estimates and our
completeness estimates using realistic extended emitters, we can
assess the most likely Schechter parameters that would lead to the
observed distribution of fluxes.
We begin by splitting
the data into three broad redshift bins covering the redshift range {\mbox{ ($2.91 \le z \le
    6.64$)}} of $\Delta\, z \approx 1$, and prepare the sample in the following ways.

\subsubsection{Completeness correction}
As introduced in Section \ref{fake_recovery} we sample the detection
completeness on a fine grid of input flux and redshift (or observed
wavelength) values with resolution $\Delta\,z = 0.01$, and $\Delta f =
0.05$ (erg$^{-1}$ cm$^{-2}$). Considering where our observed data
    lie on this grid of completeness estimates, we can then correct the
    number of objects observed at each $z$-$f$ combination to account for
    the completeness of the survey. It is these completeness-corrected
  counts that we propagate to the maximum likelihood analysis applying
the cuts described below. For a single object which
  falls at a flux brighter than the grid of combinations tested we interpolate between
  the completeness at the brightest flux tested at this redshift ($> 80\%$ at $-16.5$ erg s$^-1$ cm$^-2$), and an assumed $100\%$
  completeness by a flux of $-16.0$ erg s$^-1$ cm$^-2$. 

As our data are deep, but covering a small volume of the Universe, our
dynamic range is modest $\approx 2.0$ dex and samples well below the knee
of the luminosity function. In order to fully exploit
the information in the dataset, we can use the number of objects
observed in the sample as a constraint on the possible Schecher
parameters. This introduces the problem of the uncertainty on
the number of objects in the sample where completeness
corrections are large. For this reason we choose to cut the sample in each
redshift bin at the $25\%$ completeness limit in flux for the
median redshift of the objects in each broad redshift bin. 

In Figure {\ref{fig:sample2}} we show the flux and
log-luminosity distributions of
objects from the {\emph{mosaic}} in the same three broad redshift
bins as used for the analysis in Section \ref{sect:vmax}. For each row
of plots the redshift range is given in the top left-hand corner and
three different histograms depict the distribution of fluxes
(left-hand column) or log-luminosities (right-hand column). For each
panel we show the total distribution of objects (including the
    completeness-corrected counts)
in a coloured hatched histogram. Overlaid on this is the distribution
of observed objects in filled blue bars. The final curtailed
samples cut at the $25\%$ completeness limit in flux ($\Delta\,f =
0.05$) for the median redshift of objects in the redshift bin is
overplotted in a bold cross-hatched black histogram. 

Overlaid on each panel are the completeness curves as a function of flux (or log luminosity) at
each redshift ($\Delta\,z = 0.01$) falling within the bin. Each
redshift is given by a different coloured line according to the
colour-map shown in the colour bar. The median redshift of objects in
each redshift range is emphasized in the completeness curves, and on
the colour bar. The effect of skylines is again clearly seen in the
recovery fraction, this time manifesting as a shift of the entire
completeness curve combined with a shallower slope towards the highest
redshift LAEs in the cube. 

\subsubsection{Maximum likelihood formalism}
We begin by assuming a Schechter function, written in log form as

\begin{equation}
 {\phi\,(L)\,{\rm{d log}}L= {\rm{ln10}} \, \phi^{*}
   \left(\frac{L}{L^{*}} \right) ^{\alpha +1} \, e^{-({L}/{L^{*}})}\,{\rm{d log}}L,} 
\label{eq:Schechter}
\end{equation}

\noindent where $\phi$$^{*}$, $L^{*}$ and
$\alpha$ are the characteristic number density,
characteristic luminosity,  and the gradient of the
faint-end slope respectively \citep{Schechter1976}.

Following the method described in \cite{Marshall83}  (and applied in \citealt{Drake2013} and
\citealt{Drake2015} to narrowband samples) we can describe the
distribution of fluxes by splitting the flux range into
bins small enough to expect no more than 1 object per bin, and
writing the likelihood of finding an object in bins {F$_i$} and no
objects in bins {F$_j$}, as Equation
\ref{eq:L} for a given Schechter function:

\begin{equation}
{ {\Lambda} = {\prod_{F_i} } \Psi(F_i) \, {\rm{dlog}}F \,e^{-\Psi(F_i) {\rm{dlog}}F} \prod_{F_j}\, e^{-\Psi(F_j) {\rm{dlog}}F},} 
\label{eq:L}
\end{equation}

\noindent where $\Psi(F_i)$ is the probability of detecting an object
with true
line flux between $F$ and $10^{{\rm{d log}}F}F$ (i.e. after correction
for bias in the total flux measurements). This simplifies to Equation \ref{eq:L2}, where {F$_k$} is the product over all bins:

\begin{equation}
{ {\Lambda} = {\prod_{F_i}  \Psi(F_i) \, {\rm{dlog}}F \,  \prod_{F_k}\, e^{-\Psi(F_k) {\rm{dlog}}F}}.}
\label{eq:L2}
\end{equation}

\noindent Since the value of $\phi^*$ directly follows from $L^*$, we minimise the
likelihood function, $S = -2 {\rm{ln}} \Lambda$ (Equation
{\ref{eq:S}}) for
$L^*$ and $\alpha$ only, re-scaling $\phi*$ for the $L^*$-$\alpha$ combination to
ensure that the total number of objects in the final sample is reproduced:

\begin{equation}
{ {S} = {-2 \sum }\, {\rm{ln}} \Psi(F_i) + 2 \int \Psi (F)\, {\rm{dlog}}F.}
\label{eq:S}
\end{equation}

\subsubsection{Maximum likelihood results}

The maximum likelihood Schechter parameters are presented in Table
{\ref{tab:ML_sch_params}} and Figure {\ref{fig:Banana}}. We derive Schechter
parameters with no prior assumptions on their values, and therefore
provide an unbiased result across each of the redshift ranges evaluated. The
most likely Schechter parameters in each redshift bin give steep
values of the faint-end slope $\alpha$, and values of $L^*$ which are
consistent with the literature thanks to the
re-normalisation of each LF to reproduce the total number of objects
in the sample.

Interestingly, we find increasingly steep values of the faint-end
slope $\alpha$ with increasing redshift, and indeed the $68\%$
confidence limits on $\alpha$ alone (determined from the extreme values within the $\Delta\,
S = 1$ contour - details in Section \ref{sect:error_anal} below) gives the shallowest value of $\alpha$ in the lowest
redshift bin as inconsistent with the two
higher redshift bins at the $1-\sigma$ level. Using the $1/$V$_{max}$ estimator
\cite{Cassata2011} found a value of $\alpha$ that was steeper in their
{\mbox{($3.00 \le z \le 4.55$)}} redshift bin than in the interval {\mbox{($1.95 \le z \le
    3.00$)}}. In their highest redshift bin at {\mbox{($4.55 \le z \le
    6.60$)}} the data were insufficient to constrain the faint-end slope, and so the
authors fixed $\alpha$ to the average value of the lower two redshift
bins in order to measure $L^*$ and $\phi^*$.  Our measurement of the
faint-end slope with MUSE gives the first ever estimate of $\alpha$ at
redshift ($5.00 \le z < 6.64$) using data is $0.5$ dex deeper in the
measurement than previous estimates down to our $25\%$ completeness limit.
We should bear in mind that our highest redshift bin is much shallower
in luminosity than the other two, as sky lines begin to severely
hamper the detection of LAEs, and although we apply the same $25\%$
completeness cut-off at each redshift, the correction varies far more
across the bin than at the lower two redshifts (correction applied for
the median redshift of the bin, $z=5.48$ in the range $5.00 \le z \le
6.64$). Therefore the measurement of $\alpha$ is a much larger
extrapolation than in the other two bins, and should be interpreted
with caution.

\subsection{\bf{Error analysis}}
\label{sect:error_anal}
We examine the 2D likelihood contours in $L^*$-$\alpha$ space in the upper panel of
Figure \ref{fig:Banana},  and show the 68\% and 95\% joint confidence
regions which correspond to $\Delta S = 2.30$ and $6.18$ for
  two free fit parameters ($L^*$ and $\alpha$). This translates
  directly to a confidence interval for the dependent quantity of the
  luminosity density, and so we take the maximum and minimum values of
    the luminosity density within the contour (which contains $68\%$
    of the probability content for the Schechter parameters, fully
    accounting for their co-variance). The same logic applies to
    provide error bars on the SFRD which translates according to
    Equation \ref{eq:Lya}.

To estimate marginal 68\% confidence interval on single parameters, we
    take the two extremes of the $\Delta S = 1$ contours. This approach
    implicitly assumes a Gaussian distribution, but is a valid
    approximation for an extended, asymmetric probability function
    such as these \citep{James06} In
    addition note that
    for the two higher redshift luminosity functions the ellipses do
    not close towards bright values of L$^*$, therefore we can only
    place lower limits on the maximum likelihood parameters. As
    $\phi^*$ is not a free parameter in the fit, but derived by
    re-scaling the shape parameters by the number of objects observed,
    the error on this number has a different meaning: it is the
    uncertainty in $\phi^*$ resulting from the errors in the other
    parameters. Therefore to find the corresponding confidence
    interval for $\phi^*$, we simply re-scale the shape parameters at
    the two extremes of each contour such that the combination
    $L^*$, $\phi^*$, $\alpha$ reproduces the observations.

Finally, we note that if (due to our loose constraints on
    L$^*$) the reader prefers to assume a fixed value of L$^*$, for example {\mbox{L$^* = 42.7$}}
    across all redshifts here, the corresponding marginal 68\%
    confidence intervals for $\alpha$ would be {\mbox{$-1.95 >\alpha >
    -2.30$}} at \mbox{$2.91\le z < 4.00$}, {\mbox{$-2.20 >\alpha > -2.50$}} at \mbox{$4.00
\le z < 5.00$}, and {\mbox{$-2.60 >\alpha > -3.30$}} at \mbox{$5.00 \le z < 6.64$.}

\section{Discussion}
\label{sect:discuss}
\subsection{Evolution of the Ly$\alpha$ luminosity function}

The degeneracy between Schechter parameters often makes it
difficult to interpret whether the luminosity function has evolved
across the redshift range ($2.91 < z < 6.64$). Moreover, as noted in the review of \cite{Dunlop13}, comparing
Schechter-function parameters, particularly in the case of a very
limited dynamic range, can actually amplify any apparent
difference between the raw data sets. Nevertheless, it is useful to place constraints on the range of
    possible Schechter parameters in a number of broad redshift bins
    to give us a handle on the nature of the population over time.

The $\Delta S = 2.3$ contour
containing $68\%$ of the probability of all
three redshifts just overlap, ruling out any
dramatic evolution in the observed Ly$\alpha$ luminosity function
across this redshift range. This is entirely consistent with
literature results from \cite{Ouchi2008} and \cite{Cassata2011}.

The first signs
of evolution in the observed Ly$\alpha$ luminosity function have been
seen between redshift slices at $5.7$ and $6.6$ from narrowband
surveys \citep{Ouchi2008}, and this falls within our highest redshift bin. Although we
have too few galaxies to construct a reliable luminosity function at
these two specific redshifts, it is noteworthy perhaps that our most
likely Schechter parameters for the ($5.00 < z < 6.64$) potentially
reflect this evolution (in addition to $\alpha$ being steeper, $L^*$
drops just as is seen in \citealt{Ouchi2008}) -- so perhaps the evolution
at the edge of our survey range is strong enough to affect our highest redshift bin even though the median redshift of our galaxies is $z=5.48$.

\subsection{LAE contribution to the SFRD}

The low-mass galaxies detected via Ly$\alpha$ emission at high
redshift obviously provide a means to help us understand
typical objects in the early Universe, and the physical
properties of these galaxies will ultimately reveal the manner in
which they may have driven the reionisation of the IGM. As an
interesting first step we derive here the contribution our LAEs make to the
cosmic star-formation rate density (SFRD) compared to measures derived
from broadband selected samples which typically detect objects of
much higher stellar masses.  

\begin{figure}
\begin{center}
 \includegraphics[width=0.48\textwidth]{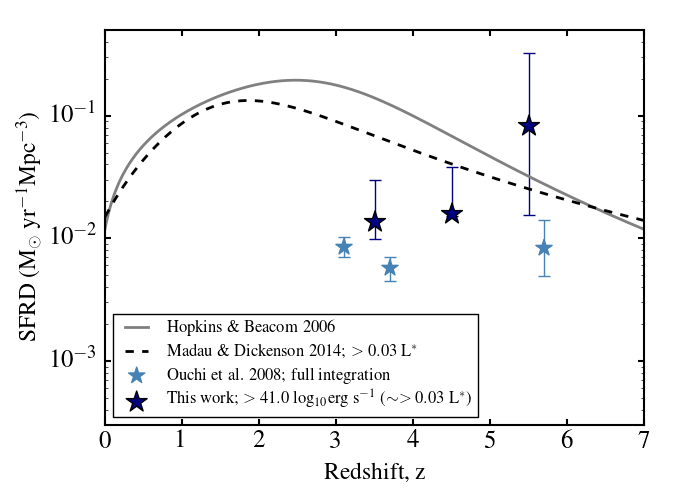} 
\caption[]{Contribution of LAEs to the cosmic SFRD in our three
  redshift bins at \mbox{$2.91\le z < 4.00$} \mbox{$4.00
\le z < 5.00$}, and \mbox{$5.00 \le z < 6.64$} integrating to log$_{10}$ L$^* = 41.0$ ($0.02$ L$^*$ in the lower
two redshift bins, and $0.03$ L$^*$ in our highest redshift bin). The
lighter blue stars show the results of Ouchi et al. (2008) for a full
integration of the Ly$\alpha$ luminosity function. The LAE results are
compared to literature studies of continuum-selected galaxies traced
by the solid and dashed lines. We
find that LAEs' contribution to the SFRD rises towards higher
redshift, although their contribution relative to that of more massive
galaxies is uncertain due to various limitations (e.g. inhomogeneous
integration limits from many different surveys, and the uncertainty in
translating Ly$\alpha$ luminosity density to an SFRD).}
\label{fig:SFRD}
\end{center}
\end{figure}

To determine our Ly$\alpha$ luminosity densities we integrate the maximum likelihood luminosity function in each of our
redshift bins to log$_{10}$ L$^* = 41.0$ ($0.02$ L$^*$ in the lower
two redshift bins, and $0.03$ L$^*$ in our highest redshift bin, shown
in the penultimate column of Table
\ref{tab:ML_sch_params}). We make the assumption that the
entirety of the Ly$\alpha$ emission is produced by star-formation,
and use Equation \ref{eq:Lya}:

\begin{equation}
{{\rm{SFR}_{Ly\alpha}}  {\rm{M_{\odot}}}  {\rm{yr}}^{-1} {\rm{Mpc}}^{-3} =  {\rm{L}_{Ly\alpha} erg s}^{-1} / 1.05 \times 10.0^{ 42.0},}
\label{eq:Lya}
\end{equation}

\noindent as in \cite{Ouchi2008}, to convert the
Ly$\alpha$ luminosity density to an SFRD. We note that
this is a very uncertain conversion, however we show in Figure \ref{fig:SFRD} our best estimates of the LAE SFRD derived from
the Ly$\alpha$ line, over-plotted on two parameterisations of the
global SFRD from $z =7$ to the present day (from \citealt{HopkinsBeacom06}
and \citealt{MadauDickenson14} which compile estimates from rest-frame UV through to
radio). These studies also faced of course the question of where to place
    the integration limit for luminosity functions drawn from the
    literature. \cite{MadauDickenson14} for example chose to use a cutoff at
    $0.03$ L$^*$ across all wavelengths in an attempt to homogenise
    the data. The limit is comparable to our own, however we note that
    by changing the integration limit of either our own luminosity
    functions, or those from the literature, one could draw very
    different conclusions as to the fraction of the total SFRD that
    LAEs are contributing.

\cite{Ouchi2008} used $858$ narrowband selected LAEs to
estimate the SFRD in three redshift slices at $z = 3.1, 3.7$ and
$5.7$ assuming a Ly$\alpha$
escape fraction $= 1$ (shown in Figure \ref{fig:SFRD} by the light blue stars). They concluded that on average LAEs contribute $\approx 20\%$ to $\approx
40\%$ of the SFRD from broadband selected surveys over the entire period.

Similarly, \cite{Cassata2011} used the VVDS spectroscopic
survey to make the same measurement using Ly$\alpha$ LFs (with
luminosities offset from their observed values according to the IGM
attenution prescription of \citealt{Fan06}). With this approach they
compared the contribution of LAEs to the SFRD as measured from LBG surveys. Only a fraction of
LBGs are also LAEs (when LAEs are defined to have a Ly$\alpha$
equivalent width greater than some cutoff) and the fraction of LAEs
amongst LBGs is known to increase towards fainter UV magnitudes. With
MUSE we probe a population of galaxies that are fainter than average
in the UV, in contrast to the majority of objects defined as LBGs. As
such, it is difficult to state what fraction of the {\it{overall}}
SFRD LAEs contribute. \cite{Cassata2011} found that the LAE-derived SFRD increases from $\approx 20\%$  at $z \approx 2.5$ to $\approx
100\%$ by $z \approx 6.0$ relative to LBG estimates. We find very similar results from our
observed luminosity functions. In our lowest redshift bin LAEs
contribute $10$ - $20\%$ of the SFRD depending on whether one compares
to the \cite{MadauDickenson14} or \cite{HopkinsBeacom06}
parameterisations, reaching $100\%$ by redshift $z \approx 6.0$. In
fact our best estimate of the SFRD at this redshift is actually
greater than the estimates from other star-formation tracers, probably
indicating the inadequacies of making a direct transformation from
Ly$\alpha$ luminosity to a star-formation rate in addition to the
    very different sample selections (the majority of previous surveys
  trace massive continuum-bright sources). It would appear that the steep
values of $\alpha$ we measure, and the steepening
    of the slope with increasing redshift easily allow the resultant
    SFRD to match or even exceed estimates from broadband selected galaxies. This means that any
    further boost in the luminosity density (such as from introducing
    an IGM attenuation correction) would act to raise LAEs'
    contribution to the broadband selected SFRD further, such that LAEs may play
    a more significant role in powering the early Universe than first thought.  

This result is not a huge surprise, as we know that the steeper the luminosity
    function, the more dramatically the Ly$\alpha$ luminosity density
    (and hence the SFRD) increases for a given
    integration limit (also discussed in section 7.3 of \citealt{Drake2013}). Thus, it follows that our highest
    redshift luminosity function produces a significantly greater SFRD
  when integrated to the same limit as the two lower redshift bins, largely driven by the steep values of $\alpha$ we
    measure. Indeed, this behaviour of the integrated luminosity
    function is one of the drivers of the need to accurately measure
    the value of $\alpha$ for the high redshift population.

\begin{figure}
\begin{center}
 \includegraphics[width=0.48\textwidth]{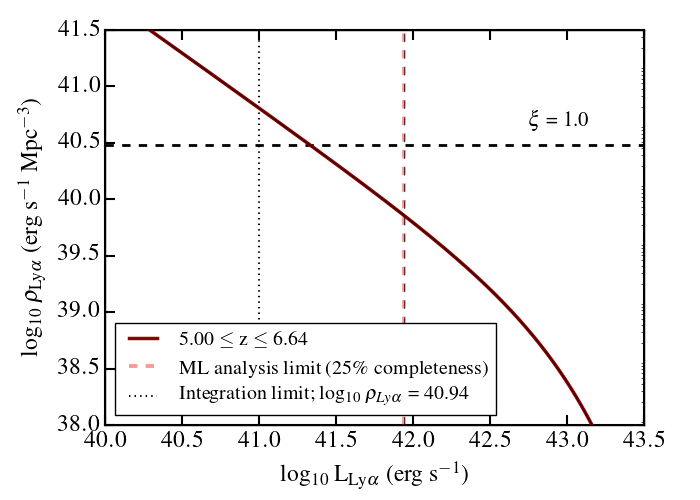} 
\caption[]{Integrating the maximum likelihood Schechter function at {\mbox{($5.00
\le z < 6.64$)}} to log L = 41.0 produces enough
ionising radiation to maintain an ionised IGM at $z \approx 6$ for the
set off assumptions described in the text.}
\label{fig:LF_reion}
\end{center}
\end{figure}

\subsection{Ly$\alpha$ luminosity density and implications for reionisation}

In fact, it is the available ionising luminosity density which is the
deciding factor in whether a given population were able to maintain an
ionised IGM. As such, some groups have attempted to compute
the critical Ly$\alpha$ luminosity that would translate to a sufficient ionising flux to maintain a
transparent IGM. To place any constraint on this value at all, it is
necessary to make various assumptions about the escape of Ly$\alpha$, the escape of Lyman continuum
(LyC) and the clumping of the IGM.

We follow the arguments laid out in \citealt{Martin08} (also \citealt{Dressler11}, 
\citealt{Henry12}, \citealt{Dressler15}),  who take fiducial values of
Ly$\alpha$ escape ($f_{esc}^{Ly\alpha} = 0.5$; \citealt{Martin08}), escape
of LyC ($f_{esc}^{LyC} = 0.1$; \citealt{Chen07}, \citealt{Shapley06}) and the clumping of the IGM ($C = 6$; \citealt{Madau99})
to determine a critical value of log$_{10}$ $\rho_{{\rm{\,Ly}}\alpha}
= 40.48$ erg s$^{-1}$ Mpc$^{-3}$  at
$z = 5.7$ (equation 5 of \citealt{Martin08}):

\begin{eqnarray}
\rho_{{\rm{Ly}}\alpha} = 3.0 \times 10^{40} {\rm{erg\,s}}^{-1}
{\rm{Mpc^{-3}}} 
\times\, C_6(1-0.1 f_{{{\rm{Lyc}},0.1}}) \nonumber \\
 \left(\frac{f_{{{\rm{Ly}}\alpha},0.5}}{ f_{{{\rm{Lyc}},0.1}}} \right)
\left(\frac{1+z}{6.7} \right)^3 \left(\frac{\Omega_{\,\beta}\, h^2_{\,70}}{0.047} \right)^2.
\end{eqnarray}

In Figure \ref{fig:LF_reion} we show the cumulative Ly$\alpha$
luminosity density, $\rho_{Ly\alpha}$ on the ordinate, against the limit
of integration on the abscissa. Using our Schechter luminosity function at redshift {\mbox{($5.00
\le z < 6.64$)}} we integrate the most likely Schechter parameters
down to log $L =41.0$ resulting in {\rev{\bf{$\rho_{Ly\alpha} = 40.94$}}}.

We need only extrapolate by $< 1$ dex beyond the $25\%$
completeness limit (the lowest luminosity galaxies included in the
maximum likelihood analysis) in order to achieve a great enough
$\rho_{Ly\alpha}$ to maintain reionisation, assuming our assumptions
are valid.


\section{Conclusions}

We have presented a large, homogeneously-selected sample of
$604$ LAEs in total from the MUSE-GTO observations of the
{\emph{HUDF}}. Using automatic detection software we build samples of
$123$ and $481$ LAEs in the {\emph{udf-10}} and
{\emph{UDF-mosaic}} fields respectively. We simulate realistic
extended LAEs based on the halo measurements of \cite{Wisotzki16} and
L17 to derive a fully-characterised LAE selection function for the
Ly$\alpha$ luminosity function. As such we compute
the deepest-ever Ly$\alpha$ luminosity function in a blank-field,
taking into account extended Ly$\alpha$ emission, and using
two different estimators to reduce the biases of a single
approach. Our main findings can be summarised as follows:\\

\begin{itemize}
\item  We find a steep faint-end slope of the Ly$\alpha$ luminosity
  function in each of our redshift bins using both the $1/$V$_{max}$-
  and maximum-likelihood estimators.

\item We see no evidence of a strong evolution in the observed luminosity
  functions between our three $68\%$ confidence regions for
  L$^*$-$\alpha$ in redshift bins at \mbox{$2.91\le z < 4.00$} \mbox{$4.00
\le z < 5.00$}, and \mbox{$5.00 \le z < 6.64$}.

\item Examining the faint-end slope $\alpha$ alone, we find an
  increase in the steepness of the luminosity function with increasing
  redshift. 

\item  LAEs contribute significantly to the cosmic SFRD,
  reaching $100\%$ of that coming from continuum-selected LBG
      galaxies by redshift $z \approx 6.0$, using the very similar integration limits and the
  Ly$\alpha$ line flux to trace star formation activity. The increase
  is partly driven by the very steep faint-end slope at \mbox{($5.00 \le z < 6.64$)}.

\item LAEs undoubtedly produce a large fraction of the ionising radiation
  required to maintain a transparent IGM at $z \approx 6.0$. Taking
  fiducial values of several key factors, the maximum likelihood
  luminosity function requires only a small
  extrapolation beyond the data ($0.8$ dex) for LAEs alone to power reionisation.\\
\end{itemize}

The ability of MUSE to capture extended Ly$\alpha$ emission around
individual high-redshift galaxies is transforming our view of the
early Universe. Now that we are an order of magnitude more sensitive to
Ly$\alpha$ line fluxes we find that faint LAEs were even more abundant in
the early Universe than previously thought. In the near
future, systematic surveys of Ly$\alpha$ line profiles from MUSE will
allow us to select galaxies which are likely to be leaking LyC
radiation, and in conjunction with simulations this will lead to a
better understanding of the way that LAEs were able to power the
reionisation of the IGM.

\begin{acknowledgements}

This work has been carried out thanks to the support of the ANR FOGHAR (ANR-13-BS05-0010-02), the OCEVU Labex (ANR-11-LABX-0060) and the A*MIDEX project (ANR- 11-IDEX- 0001-02) funded
by the ``Investissements d'avenir'' French government program managed
by the ANR. ABD would like to acknowledge the following people; Dan Smith, Richard Parker, Phil James, Helen Jermak, Rob Barnsley, Neil Clay, Daniel Harman,
   Clare Ivory, David Eden, David Lagattuta, David Carton and my
   family. JS acknowledges the European Research Council under the European
Unions Seventh Framework Programme (FP7/2007-
2013) / ERC Grant agreement 278594-GasAroundGalaxies. JR acknowledges
support from the ERC starting grant 336736-CALENDS. 
TG is grateful to the LABEX Lyon Institute of Origins (ANR-10-LABX-0066) of the Université de Lyon for its financial support within the program "Investissements d'Avenir" (ANR-11-IDEX-0007) of the French government operated by the National Research Agency (ANR).
\end{acknowledgements}

\bibliographystyle{aa} 

\begin{appendix}
\section{Vmax results}
\begin{table*}
\caption{Differential Ly$\alpha$ luminosity function in bins of
  $\Delta$ log10 L= 0.4 using the $1/V_{max}$ estimator. Errors
      quoted on values of $\phi$ are $1\sigma$ assuming Poissonian statistics.}  
\label{tab:vmax} 
\centering         
\renewcommand{\arraystretch}{1.2}         
\begin{tabular}{c c c c}   
\hline \hline 
\multicolumn{4}{c}{\bf{Redshift Bin (2.92 $\le z < $   4.00) }}\\
 Bin log$_{10}$ ($L$) [erg s$^{-1}$]& log$_{10}$ $L$$_{{\rm{median}}}$
[ergs$^{-1}$] & $\phi$ [(dlog$_{10}$ $L$)$^{-1}$ Mpc$^{-3}$]  & No. objects \\ 
\hline 
 $  41.00 <{\bf{ 41.200}} <  41.40$ & $ 41.309$ & $0.02086\pm 0.00467$ & $     20$\\ 
 $  41.40 <{\bf{ 41.600}} <  41.80$ & $ 41.633$ & $0.03846\pm 0.00351$ & $    120$\\ 
 $  41.80 <{\bf{ 42.000}} <  42.20$ & $ 41.967$ & $0.01125\pm 0.00129$ & $     76$\\ 
 $  42.20 <{\bf{ 42.400}} <  42.60$ & $ 42.316$ & $0.00374\pm 0.00082$ & $     21$\\ 
 $  42.60 <{\bf{ 42.800}} <  43.00$ & $ 42.807$ & $0.00013\pm 0.00009$ & $      2$\\ 
\hline \hline
\multicolumn{4}{c}{\bf{Redshift Bin (4.00 $\le z < $   5.00)}}\\
Bin log$_{10}$ ($L$) [erg s$^{-1}$]& log$_{10}$ $L$$_{{\rm{median}}}$
[ergs$^{-1}$] & $\phi$ [(dlog$_{10}$ $L$)$^{-1}$ Mpc$^{-3}$]  & No. objects \\ 
\hline 
 $  41.00 <{\bf{ 41.200}} <  41.40$ & $ 41.301$ & $0.01871\pm 0.00454$ & $     17$\\ 
 $  41.40 <{\bf{ 41.600}} <  41.80$ & $ 41.660$ & $0.02489\pm 0.00284$ & $     77$\\ 
 $  41.80 <{\bf{ 42.000}} <  42.20$ & $ 41.968$ & $0.01137\pm 0.00138$ & $     68$\\ 
 $  42.20 <{\bf{ 42.400}} <  42.60$ & $ 42.375$ & $0.00249\pm 0.00052$ & $     23$\\ 
 $  42.60 <{\bf{ 42.800}} <  43.00$ & $ 42.785$ & $0.00145\pm 0.00065$ & $      5$\\ 
 $  43.00 <{\bf{ 43.200}} <  43.40$ & $ 43.071$ & $0.00008\pm 0.00008$ & $      1$\\
\hline \hline 
\multicolumn{4}{c}{\bf{Redshift Bin (5.00 $\le z < $   6.64)}}\\
Bin log$_{10}$ ($L$) [erg s$^{-1}$]& log$_{10}$ $L$$_{{\rm{median}}}$
[ergs$^{-1}$] & $\phi$ [(dlog$_{10}$ $L$)$^{-1}$ Mpc$^{-3}$]  & No. objects \\ 
\hline 
 $  41.00 <{\bf{ 41.200}} <  41.40$ & $ 41.235$ & $ 0.0049\pm  0.0028$ & $      3$\\ 
 $  41.40 <{\bf{ 41.600}} <  41.80$ & $ 41.664$ & $ 0.0077\pm  0.0016$ & $     22$\\ 
 $  41.80 <{\bf{ 42.000}} <  42.20$ & $ 42.000$ & $ 0.0073\pm  0.0012$ & $     36$\\ 
 $  42.20 <{\bf{ 42.400}} <  42.60$ & $ 42.321$ & $ 0.0034\pm  0.0007$ & $     26$\\ 
 $  42.60 <{\bf{ 42.800}} <  43.00$ & $ 42.744$ & $ 0.0008\pm  0.0003$ & $      7$\\ 
 $  43.00 <{\bf{ 43.200}} <  43.40$ & $ 43.194$ & $ 0.0001\pm  0.0001$ & $      1$\\  
\hline \hline
\multicolumn{4}{c}{\bf{Global Sample, Redshift (2.92 $\le z < $ 6.64)}}  \\ 
Bin log$_{10}$ ($L$) [erg s$^{-1}$]& log$_{10}$ $L$$_{{\rm{median}}}$
[ergs$^{-1}$] & $\phi$ [(dlog$_{10}$ $L$)$^{-1}$ Mpc$^{-3}$]  & No. objects \\ 
\hline 
 $  41.00 <{\bf{ 41.200}} <  41.40$ & $ 41.303$ & $0.02679\pm 0.00424$ & $     40$\\ 
 $  41.40 <{\bf{ 41.600}} <  41.80$ & $ 41.643$ & $0.03773\pm 0.00255$ & $    219$\\ 
 $  41.80 <{\bf{ 42.000}} <  42.20$ & $ 41.971$ & $0.01377\pm 0.00103$ & $    180$\\ 
 $  42.20 <{\bf{ 42.400}} <  42.60$ & $ 42.333$ & $0.00385\pm 0.00046$ & $     70$\\ 
 $  42.60 <{\bf{ 42.800}} <  43.00$ & $ 42.766$ & $0.00079\pm 0.00021$ & $     14$\\ 
 $  43.00 <{\bf{ 43.200}} <  43.40$ & $ 43.133$ & $0.00004\pm 0.00003$ & $      2$\\ 
\hline 
\end{tabular}
\end{table*}

\end{appendix}

\end{document}